\documentclass[aps,preprint,floatfix,nofootinbib,showpacs]{revtex4-1}
\pdfoutput=1
\usepackage{graphicx,color}
\usepackage{hyperref}
\usepackage{amsmath}
\usepackage{mathrsfs}
\usepackage{dcolumn}
\usepackage{subfigure}
\usepackage{multirow}
\begin{document}
\title{Vector Boson Fusion versus Gluon Fusion}
\renewcommand{\thefootnote}{\arabic{footnote}}

\author{
Chen-Hsun Chan$^1$, Kingman Cheung$^{1,2,3}$, Yi-Lun Chung$^1$, 
and Pai-Hsien Hsu$^1$}
\affiliation{
$^1$ Department of Physics, National Tsing Hua University,
Hsinchu 300, Taiwan \\
$^2$  Physics Division, National Center for Theoretical Sciences, Hsinchu,
Taiwan \\
$^3$ Division of Quantum Phases and Devices, School of Physics, 
Konkuk University, Seoul 143-701, Republic of Korea \\
}
\date{\today}

\begin{abstract}
Vector-boson fusion (VBF) is a clean probe of the electroweak-symmetry
breaking (EWSB), which inevitably suffers from some level of 
contamination due to the gluon fusion (ggF). In addition to the jet variables
used in the current experimental analysis, we analyze a few more 
jet-shape variables defined by the girth and integrated jet-shape.
Taking $H\to W W^* \to e \nu \mu \nu$ and $H \to \gamma\gamma$ 
as examples, we perform the analysis with a new technique of 2-step
boosted-decision-tree method, which significantly reduces the
contamination of the ggF in the VBF sample, thus, providing a clean
environment in probing the EWSB sector.
\end{abstract}

\maketitle

\section{Introduction}
The origin of mass is one of the most fundamental questions for our existence.
Particle physics explains the origin of mass by the electroweak
symmetry breaking (EWSB).  Before the electroweak symmetry is broken
the whole Universe is filled up with a Higgs field and every particle
is massless.  When this Higgs field develops a vacuum expectation
value (VEV), a particular direction in the field space is chosen and
the symmetry is broken. Particles then acquire masses
proportional to the VEV of the Higgs field. 

The discovery of the Higgs boson 
at the Large Hadron Collider (LHC) in 2012 \cite{higgs}
was a remarkable evidence of the EWSB and its properties help us
to fully understand the nature of the EWSB.
The long-sought standard model (SM) Higgs boson
was proposed more than 50 years ago, which breaks the electroweak
symmetry in order to give masses to gauge bosons and fermions. If the
discovered boson is really the SM Higgs boson or something similar, the
investigation of its properties would give a lot of information about
the EWSB.

The measurements of the properties of the Higgs boson, including mass,
total width, production cross sections, and branching ratios will 
give us a lot of information on its
gauge and Yukawa couplings, thus indirectly the details inside the
EWSB sector, which could be as complicated as one can imagine. The current
dominant production mechanism of the Higgs boson is the gluon fusion (ggF),
followed by a small fraction by vector-boson fusion (VBF).  Although the
ggF could provide useful information on the top-Yukawa coupling, the
VBF is the ultimate testing ground for probing the EWSB section, because
the longitudinal component of the $W$ and $Z$ bosons originate from the
EWSB sector itself. 

The approach of isolating the VBF from ggF relies on the properties of
the jets involved in the process and a few techniques were developed 
two decades ago, namely, forward-jet tagging
\cite{forward} and central-jet vetoing \cite{veto}. The two
accompanying jets carry most of the jet energy of the incoming quark
partons, and thus they are very energetic and very forward. One can
also make use of the wide rapidity gap between those two jets \cite{rap}.
On the other hand, the jets involved in the ggF come directly from the
QCD radiation. 
Naively, we would expect a very rich event sample of VBF from the 
experimental data with all the sophisticated jet selection cuts.
Nevertheless, with much improved accuracy in the 
N$^3$LO calculation of ggF \cite{n3LO}
the level of ggF in such selection is indeed not negligible but a
substantial fraction of the VBF$+$ggF sample. We shall use
the word ``contamination'' of the VBF sample to denote the
fraction of ggF in the VBF$+$ggF sample.
\footnote
{In this study, although we generate the VBF Monte-Carlo sample and ggF sample 
separately, we shall keep using ``contamination'' to denote the fraction
of ggF in the sum VBF$+$ggF events.
}
Thus, the ``contamination'' of the VBF sample due to
ggF is defined by
\[
  \frac{ \rm ggF} { {\rm VBF} + {\rm ggF} } \;.
\]
It stands at a level about 25\% in the current experimental studies
\cite{atlas-vbf-ww,atlas-vbf-gg}.
The purer the VBF sample, the better one can probe the EWSB sector. 
The current experimental status of discriminating the VBF from ggF was based
on a set of jet kinematical variables ($M_{jj}$, $\Delta \eta_{jj}$, ...),
a set of jet-shape variables, and those kinematic variables 
depending on the decay channel of the Higgs boson. 
A standard boosted-decision-tree (BDT) approach was employed to achieve
the current purity of the VBF sample and to reduce the contamination of the 
ggF.  
Note that the purity of the VBF is defined here as
\[
   \frac{ \rm VBF} { {\rm VBF} + {\rm ggF} + \mbox{other SM background} } \;.
\]
In this study, we employ a 2-step BDT analysis to further reduce the
contamination by ggF, thus a purer VBF sample is achieved without
significant loss in event rates.  This is the main result of this
work.  We illustrate our analysis for the decay channels of
$H\to W W^* \to e  \nu \mu \nu$ and $H \to \gamma\gamma$.

The organization is as follows. In the next section, we describe the
Monte-Carlo simulations, and in Sec. III procedures in the BDT analysis.
We present the results in Sec. IV and conclude in Sec. V.

\section{Event samples preparation}

In order to compare directly with the current status on purity of 
VBF samples of ATLAS \cite{atlas-vbf-ww,atlas-vbf-gg}, we 
follow their preparation of event samples as closely as possible.
We simulate the event samples for Higgs boson production 
including those via VBF and ggF using the P{\scriptsize OWHEG} 
\cite{powheg,powhegVBF, powhegggF}
generator at next-to-leading-order (NLO), with 
input parton distribution functions (PDFs) \scriptsize CT\small10 \cite{CT10},
and
the mass and width of the Higgs taken at $m_H\,=\,125$GeV and
$\Gamma_H\,=\,4.07$MeV. The Higgs boson samples are normalized to the
cross sections given in the ATLAS analysis for $13\,\mathrm{TeV}$. 
Note that for $H\to\gamma\gamma$ a parton-level cut 
$105\leq m_{\gamma\gamma}\leq 160 \,\mathrm{GeV}$ (Higgs window) is applied. 

All Higgs boson events are then showered and decayed into either
$WW+\mathrm{jets}$ or $\gamma\gamma+\mathrm{jets}$ by P{\scriptsize
YTHIA} {\small 8} \cite{pythia8}
and passed to D{\scriptsize ELPHES} \cite{delphes3} 
\footnote{
Version 3 is used here with the anti-$k_T$ jet algorithm using
$\Delta R =0.4$ and $p_{T_j}^{\rm min} = 20$ GeV and the
$b$-tagging efficiency is given by
$0.80 \tanh (0.003\, p_T ) \frac{30}{1 + 0.086\, p_T}$, where $p_T$ is given
in GeV.
}
for detector-level
simulation. Note that for the channel $H\to WW^*$ each of the $W$ bosons 
further decays into a charged lepton and a neutrino. 
Note that the charged-lepton flavors from the $W$ boson pair are 
required to be different, i.e, $e^+ \mu^-$ or $e^- \mu^+$. 
Table \ref{tab:MC_HWW} summarizes the event generators and the cross sections 
for each process.

 \begin{table}[h]
    \begin{tabular}{l@{\qquad}l@{\qquad}r@{.}l@{\qquad\quad}l}
    \hline\hline
         Process & MC generator & \multicolumn{2}{l}{$\sigma\cdot\mathcal{B}\, \left(\mathrm{pb}\right)$} & Number of Events \\
    \hline
    VBF &P\scriptsize OWHEG \small +P\scriptsize YTHIA \small 8 &0&0232  & 
\hspace{1.09ex}553240\\
    ggF &P\scriptsize OWHEG \small +P\scriptsize YTHIA \small 8 &0&297 & 1936340\\
    $t\overline{t}$ &M\scriptsize AD\small G\scriptsize RAPH\small 5\_\scriptsize A\small MC@NLO +P\scriptsize YTHIA \small 8 &22&6 & 3319440\\
    $WW$ &P\scriptsize OWHEG \small +P\scriptsize YTHIA \small 8 &3&10 & 
3319440\\
   \hline\hline
    \end{tabular}
    \caption{Monte Carlo generators, cross sections and the generated number 
of events (non-normalized) used to model each signal and background 
process in $WW$ decay channel at $\sqrt{s}=13\,TeV$ }
    \label{tab:MC_HWW}
    \end{table}

In the $WW$ decay channel, we consider two main backgrounds: the
SM $t\bar{t}$ and $WW$ production. The $t\bar{t}$ events are
generated at NLO using the M{\scriptsize AD}{\small G}{\scriptsize
RAPH}\small 5\_{\scriptsize A}{\small MC@NLO} 
(version 2.4.3)~\cite{madgraph}, while the 
$WW$ events are generated with P{\scriptsize OWHEG} at NLO \cite{WW}. After
then, the $t\bar{t}$ and $WW$ events are showered and each top quark
decays into $b+W$ with P{\scriptsize YTHIA} {\small 8} \cite{pythia8}. 
The $W$ bosons further decay into $\ell+\nu$, and the flavors of two
charged leptons in each event are required to be different. 
Events are then passed into D{\scriptsize ELPHES} for detector simulations. 
The event generators, cross sections, and the generated number of events
for these backgrounds are also tabulated in Table \ref{tab:MC_HWW}.

In the diphoton channel, we only consider one source of
background: $\gamma\gamma+jj$, which are generated at leading-order (LO) 
using the M\scriptsize AD\small G\scriptsize RAPH\small 5\_\scriptsize A\small
MC@NLO\cite{madgraph}. Each of the jet in $\gamma\gamma+jj$ events is
then showered into multi-jets with P\scriptsize YTHIA \small6 \cite{pythia6}. 
Finally, events are passed into D{\scriptsize ELPHES} for detector 
simulations. 
The event generators, cross sections, and the generated number of events
for the backgrounds in 
the diphoton decay channel are also listed in Table \ref{tab:MC_Haa}.
\footnote
{ 
The background events of $jj\gamma\gamma$ are generated with a set of basic
cuts: $p_{T_\gamma} > 20$ GeV, $|\eta_\gamma| < 2.5$,
$p_{T_j} > 20$ GeV, and $|\eta_j | < 5$ in the generator 
level to avoid the divergence.}

 \begin{table}[h]
    \begin{tabular}{l@{\qquad}l@{\qquad}r@{.}l@{\qquad\quad}l}
    \hline\hline
         Process & Generator  &  \multicolumn{2}{l}{$\sigma\cdot\mathcal{B}\,\left(\mathrm{pb}\right)$} & Number of Events \\
    \hline
    VBF &P\scriptsize OWHEG \small +P\scriptsize YTHIA \small 8 & 0&862 &
\hspace{1.09ex}200000\\
    ggF &P\scriptsize OWHEG \small +P\scriptsize YTHIA \small 8 & 11&1 & 
\hspace{1.09ex}800000\\
    $\gamma\gamma$+jj &M\scriptsize AD\small G\scriptsize RAPH\small 5\_\scriptsize A\small MC@NLO +P\scriptsize YTHIA \small 6 & 4&12 & 2000000\\
   \hline\hline
    \end{tabular}
    \caption{Monte Carlo generators, cross sections, and the generated number 
of events (non-normalized) used to model each signal and background process in diphoton decay channel at $\sqrt{s}=13\,TeV$ }
    \label{tab:MC_Haa}
    \end{table}

\section{Methods in Boosted Decision Trees (BDT)}

The dedicated event samples will undergo a series of analysis tools or
methods, including preselection cuts and boosted decision tree (BDT) 
\cite{BDT},
in order to enhance the purity of the VBF among the Higgs signals 
and backgrounds.
In general, each signal and background event has to first pass a 
set of kinematic
preselection cuts, and then is further selected according to the BDT
output. 
In each decay channel, we present four different methods of BDT,
including the standard BDT, which mainly follows the method in ATLAS
so that  we can make directly comparison to the other three new methods
of BDT.
Tables \ref{tab:BDT_HWW} and \ref{tab:BDT_Haa} summarize the procedures 
for $H\to WW^*$ and $H\to\gamma\gamma$, respectively.
The details are described in the following two subsections.

\begin{table}[b!]
\begin{tabular}{lr}
\hline
Parameter    & value \\
\hline
 NTrees (Number of trees in the forest)  &  1000 \\
 Shrinkage  & 0.1  \\
 nCuts (number of steps during node cut optimization)  & 20 \\
  MaxDepth (Max depth of the decision tree allowed) &  2 \\
\hline
\end{tabular}
\caption{\label{bdt-inputs}
The BDT parameters that are used in various BDT runs, except for 
the 11-variable BDT and the step-2 BDT used in $H \to W W^*$ channel 
that used NTrees $= 800$ to avoid over-training. 
The event rates stay the same with the change in NTrees.
}
\end{table}

We used the Gradient BDT with the BDT parameters given 
in Table~\ref{bdt-inputs}.
We have varied a few slightly different settings, but the outputs
do not have significant changes.
The BDT is trained after the preselection cuts to improve the 
statistics of simulated samples used in the training.  The variables can
be ranked by their {\it rankings} in the training. The BDT output score 
is defined in the range of $-1$ to $1$, with signal-like events having 
a score close to 1 and background-like events a score close to $-1$.

\subsection{$H\to WW^*\to e\nu\mu\nu$}
The event samples for the VBF $H\to WW^*$ signal, ggF, and the SM backgrounds
have to pass the preselection
cuts which were given in the current ATLAS analysis for the SM
Higgs boson decaying into $WW^*$ in the different lepton-flavor
category, which are described as follows:
\begin{enumerate}
\item $N_j \geq 2$;
\item $p_{\mathrm{T}}^{j} > 25 \, \mathrm{GeV}\, 
\left(\left|\eta^{j}\right|<2.4\right)$ and $p_{\mathrm{T}}^{j} > 30 \, 
\mathrm{GeV}\, \left(2.4<\left|\eta^{j}\right|<4.4\right)$;
\item  $p_T^{\ell 1}>25\,\mathrm{GeV}$ and $p_T^{\ell 2}>15\,\mathrm{GeV}$;
\item $m_{\ell\ell}>10\,\mathrm{GeV}$, where $m_{\ell\ell}$ is the 
invariant mass of two leading leptons;
\item $N_b = 0$;
\item Outside-lepton veto (OLV), and central-jet veto (CJV) 
  \cite{atlas-vbf-ww}
\end{enumerate}

\begin{table}[th!]
    \begin{tabular}{l@{\qquad}l@{\quad}l@{\quad}l@{\quad}l}
    \hline\hline
        Objective & Standard BDT & 11-Var BDT & 7-Var BDT & 2-step BDT  \\
    \hline
    \multirow{1}{*}{Preselection } & \multicolumn{4}{c}{$N_j \geq 2$, $N_b=0$,}\\
                                                   & \multicolumn{4}{c}{$p_{\mathrm{T}}^{j} > 25 \, \mathrm{GeV}\, \left(\left|\eta^{j}\right|<2.4\right)$ \& $p_{\mathrm{T}}^{j} > 30 \, \mathrm{GeV}\, \left(2.4<\left|\eta^{j}\right|<4.4\right)$,}\\
                                                   & \multicolumn{4}{c}{$p_T^{\ell 1}>25\,\mathrm{GeV}$, $p_T^{\ell 2}>15\,\mathrm{GeV}$,}\\
                                                    & \multicolumn{4}{c}{$m_{\ell\ell}>10\,\mathrm{GeV}$,}\\
                                                   & \multicolumn{4}{c}{OLV, CJV}\\
                                                    \hline
     $1^{\mathrm{st}}$ step &&&&\\
    \quad Signal sample & VBF & VBF & VBF & VBF\\
    \quad Bkg. sample & ggF \& $t\bar{t}$ \& $WW$&ggF \& $t\bar{t}$ \& $WW$ &ggF \& $t\bar{t}$ \& $WW$& $t\bar{t}$ \& $WW$\\
    \multirow{1}{*}{\quad BDT inputs}  & $m_{jj} $, $\Delta\eta_{jj}$, $p_{\mathrm{T}}^{\mathrm{sum}}$,& $m_{jj} $, $\Delta\eta_{jj}$, $p_{\mathrm{T}}^{\mathrm{sum}}$, & $m_{jj}$, $\Delta \eta_{jj}$,& $m_{jj} $, $\Delta\eta_{jj}$, $p_{\mathrm{T}}^{\mathrm{sum}}$, \\
                                  & $\sum m_{\ell_j}$, $\sum C_\ell$,& $\sum m_{\ell_j}$, $\sum C_\ell$, & $p_{\mathrm{T}}^{\mathrm{sum}}$, $\sum m_{\ell j}$,& $\sum m_{\ell_j}$, $\sum C_\ell$,\\
                         & $m_{\ell\ell}$, $\Delta\phi_{\ell\ell}$, $m_{\mathrm{T}}$& $m_{\ell\ell}$, $\Delta\phi_{\ell\ell}$, $m_{\mathrm{T}}$ & $\sum g_{j}$, $\Psi_{\mathrm{c}}$, $\Psi_{\mathrm{s}}$& $m_{\ell\ell}$, $\Delta\phi_{\ell\ell}$, $m_{\mathrm{T}}$\\
                               & & $\sum g_{j}$, $\Psi_{\mathrm{c}}$, $\Psi_{\mathrm{s}}$& &\\
                                                                 \hline
    $2^{\mathrm{nd}}$ step &&&&\\
   \quad Signal sample  &  \quad - & \quad - & \quad - & VBF \\
\quad Bkg. sample  & \quad - & \quad - & \quad - & ggF\\
     \multirow{1}{*}{\quad BDT inputs}  & \multirow{3}{*}{\quad -} & \multirow{3}{*}{\quad -} & \multirow{3}{*}{\quad -} & $m_{jj}$, $\Delta \eta_{jj}$,  \\
                           & & & &$p_{\mathrm{T}}^{\mathrm{sum}}$, $\sum m_{\ell j}$,\\
                           & & & & $\sum g_{j}$, $\Psi_{\mathrm{c}}$, $\Psi_{\mathrm{s}}$, \\
   \hline\hline
    \end{tabular}
    \caption{Summary of each analytic method for $H\to WW^*$  }
    \label{tab:BDT_HWW}
    \end{table}

\begin{figure}[th!]
\centering
        \includegraphics[scale=0.35]{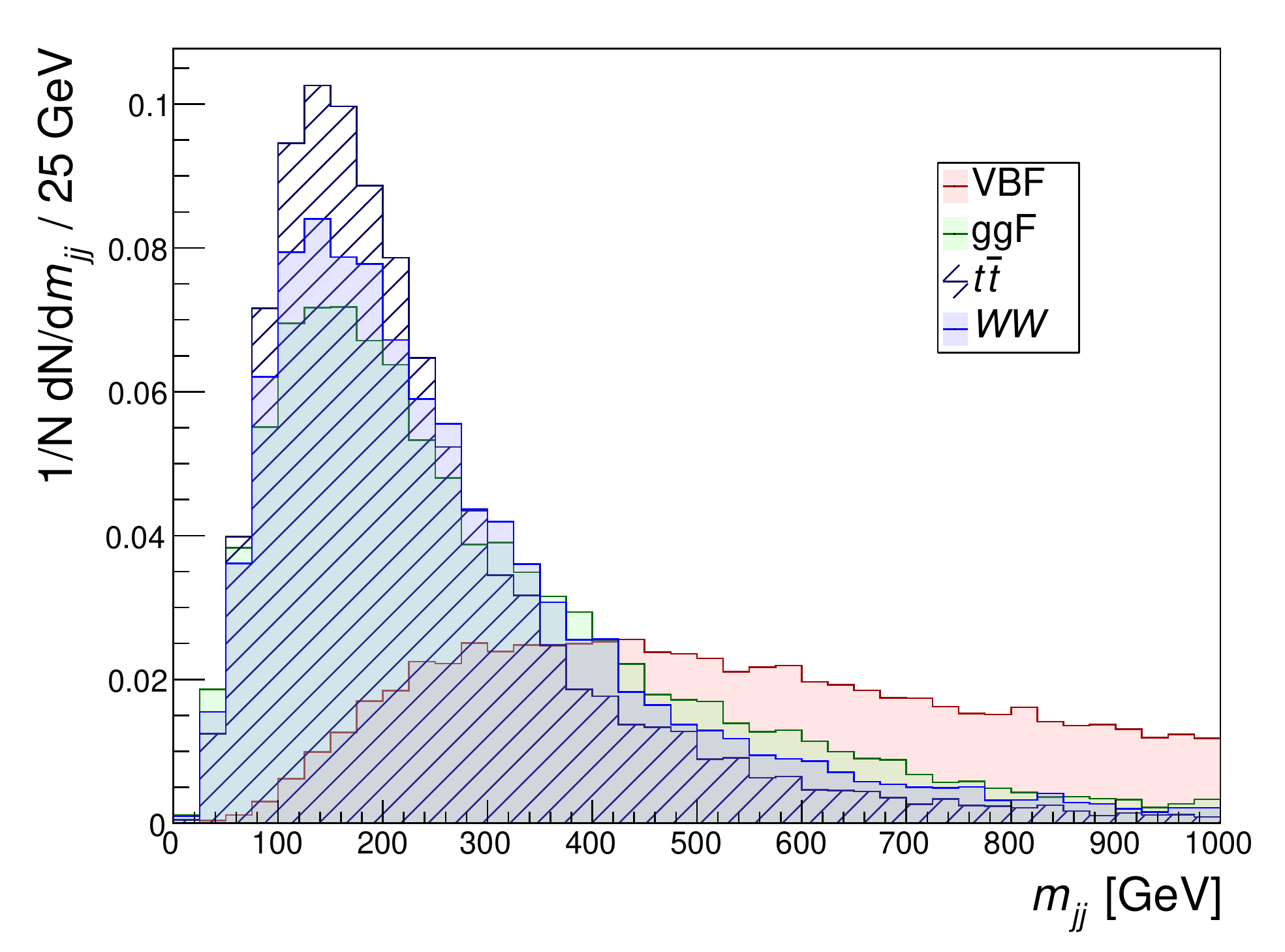}
        \includegraphics[scale=0.35]{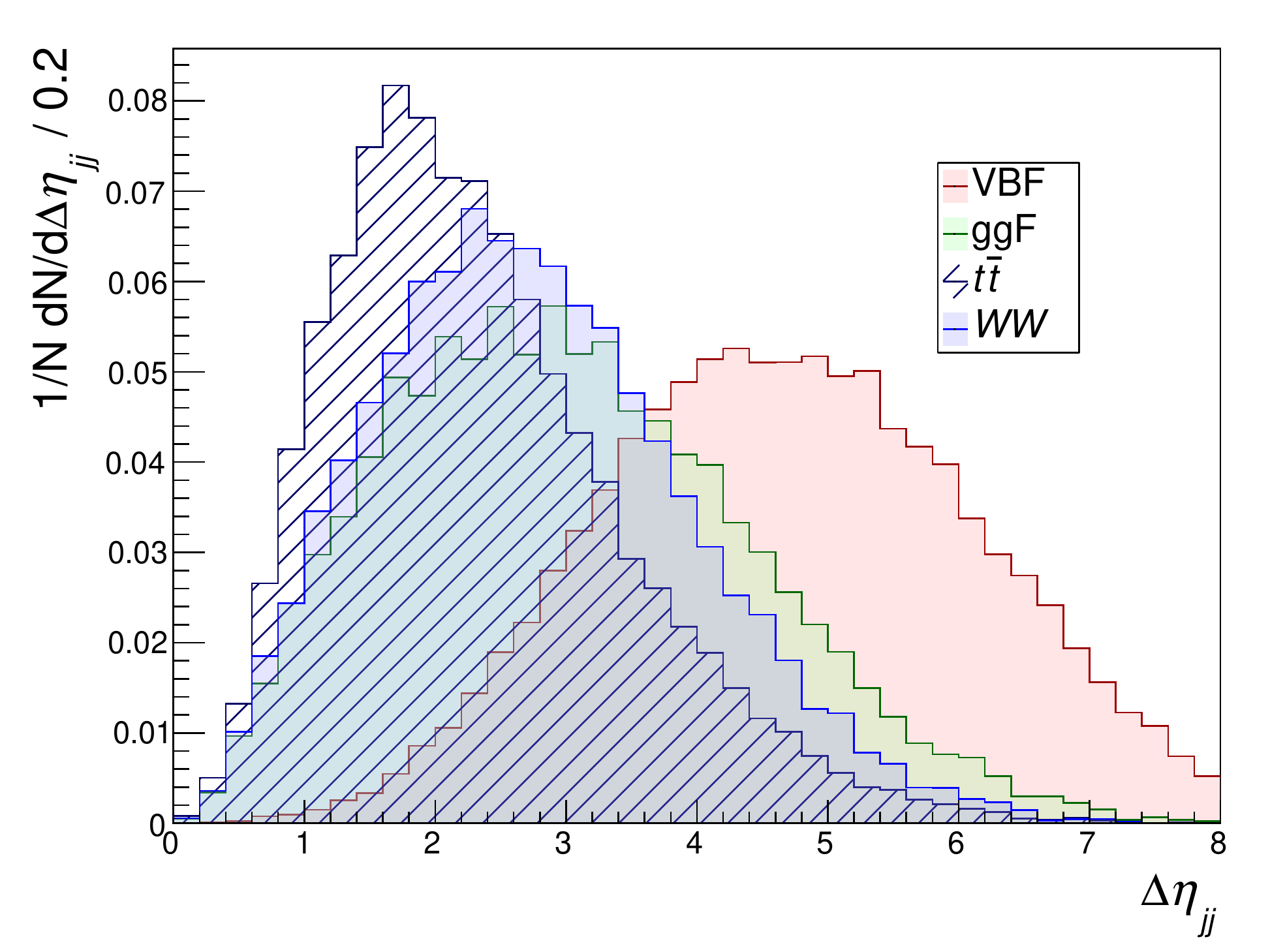}
        \includegraphics[scale=0.35]{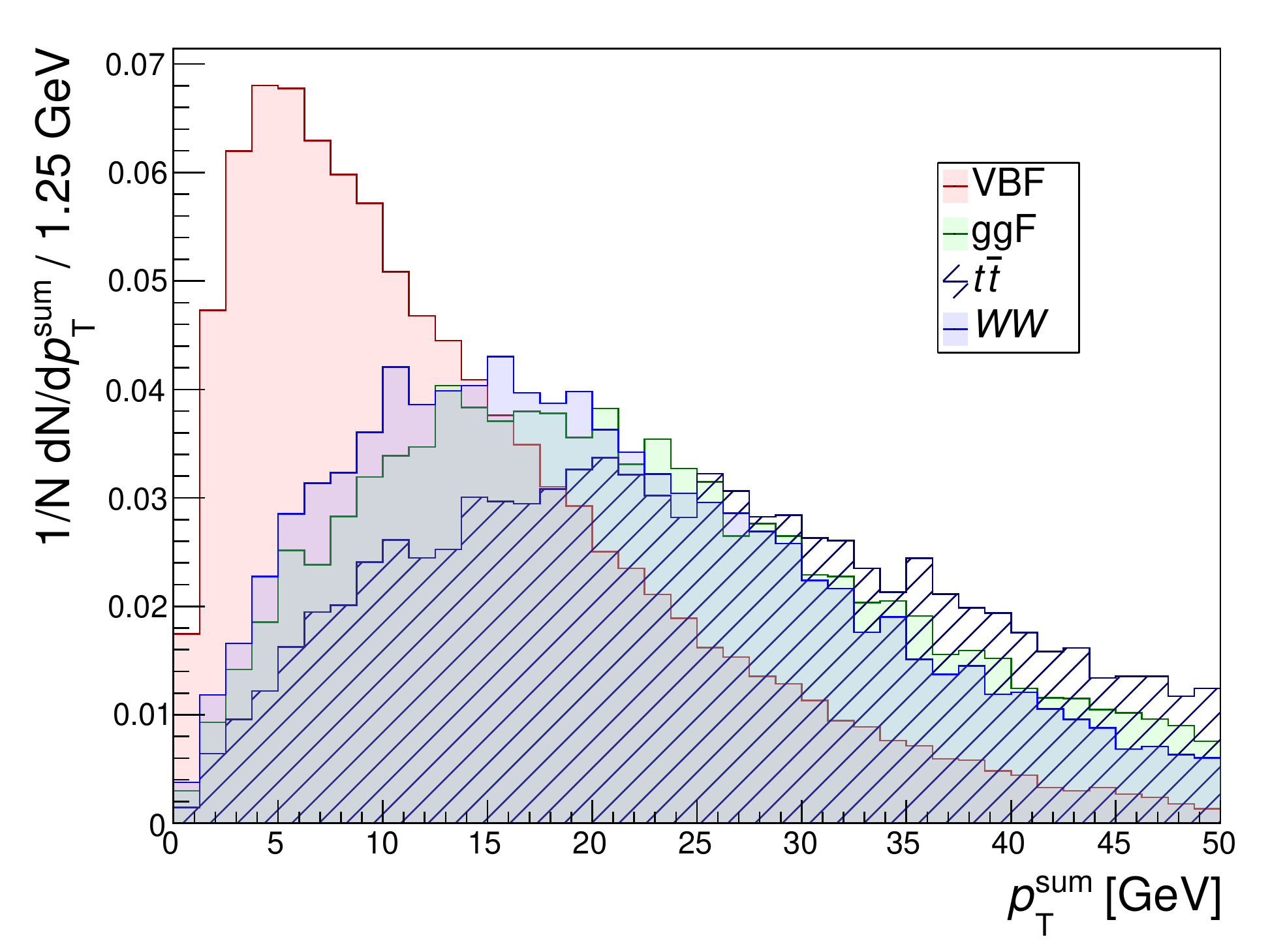}
        \includegraphics[scale=0.35]{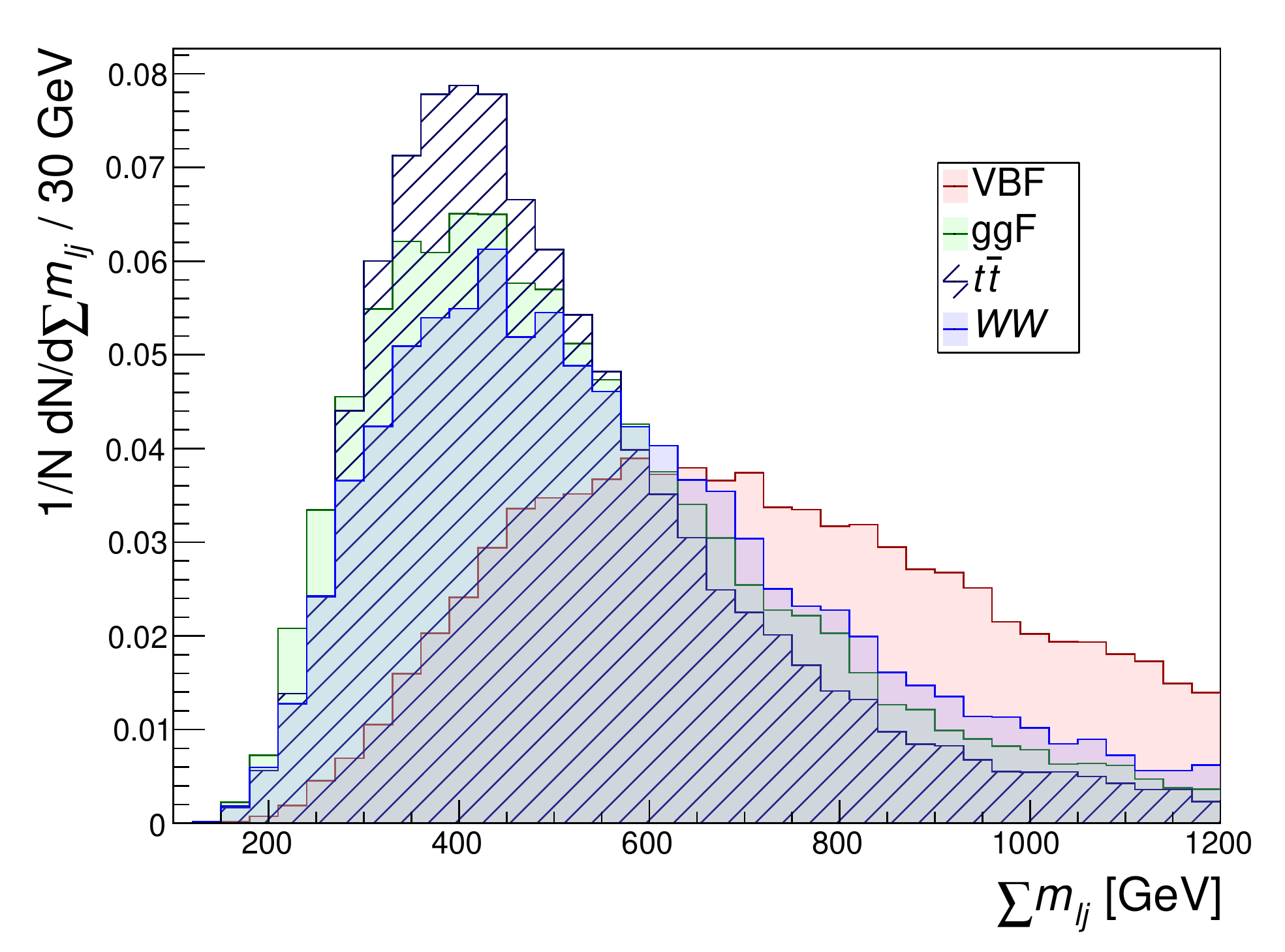}
        \includegraphics[scale=0.35]{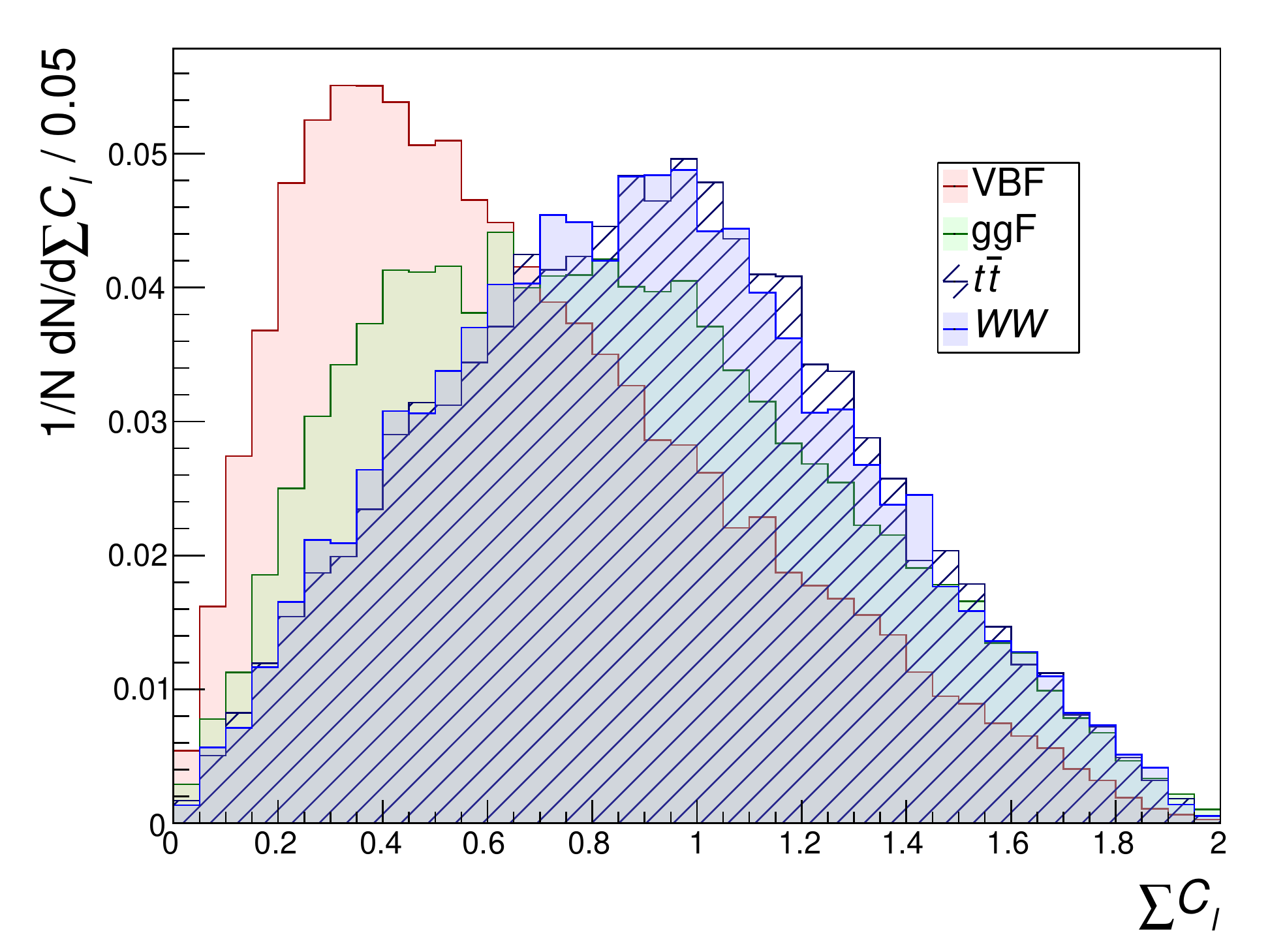}
        \includegraphics[scale=0.35]{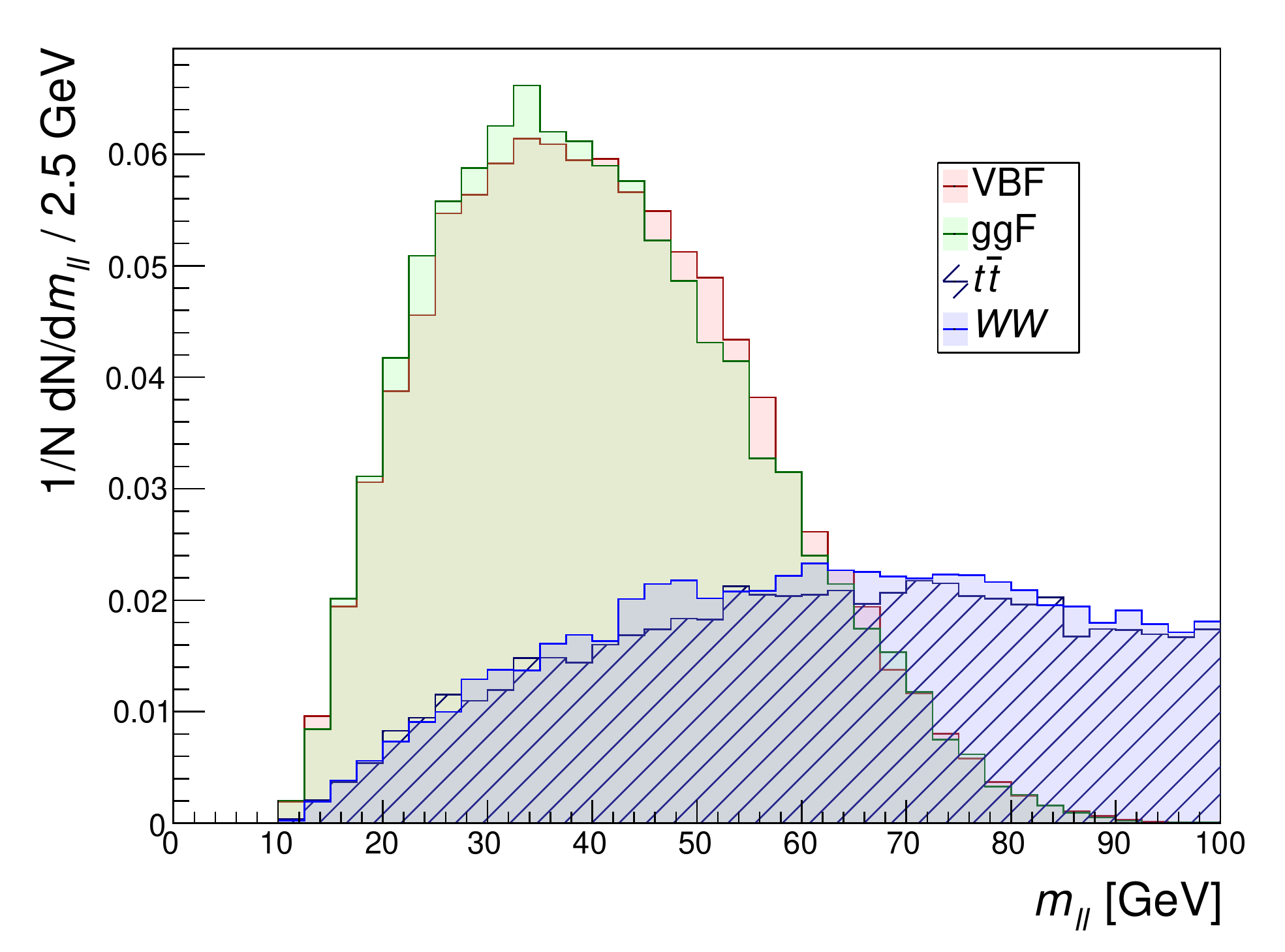}
        \includegraphics[scale=0.35]{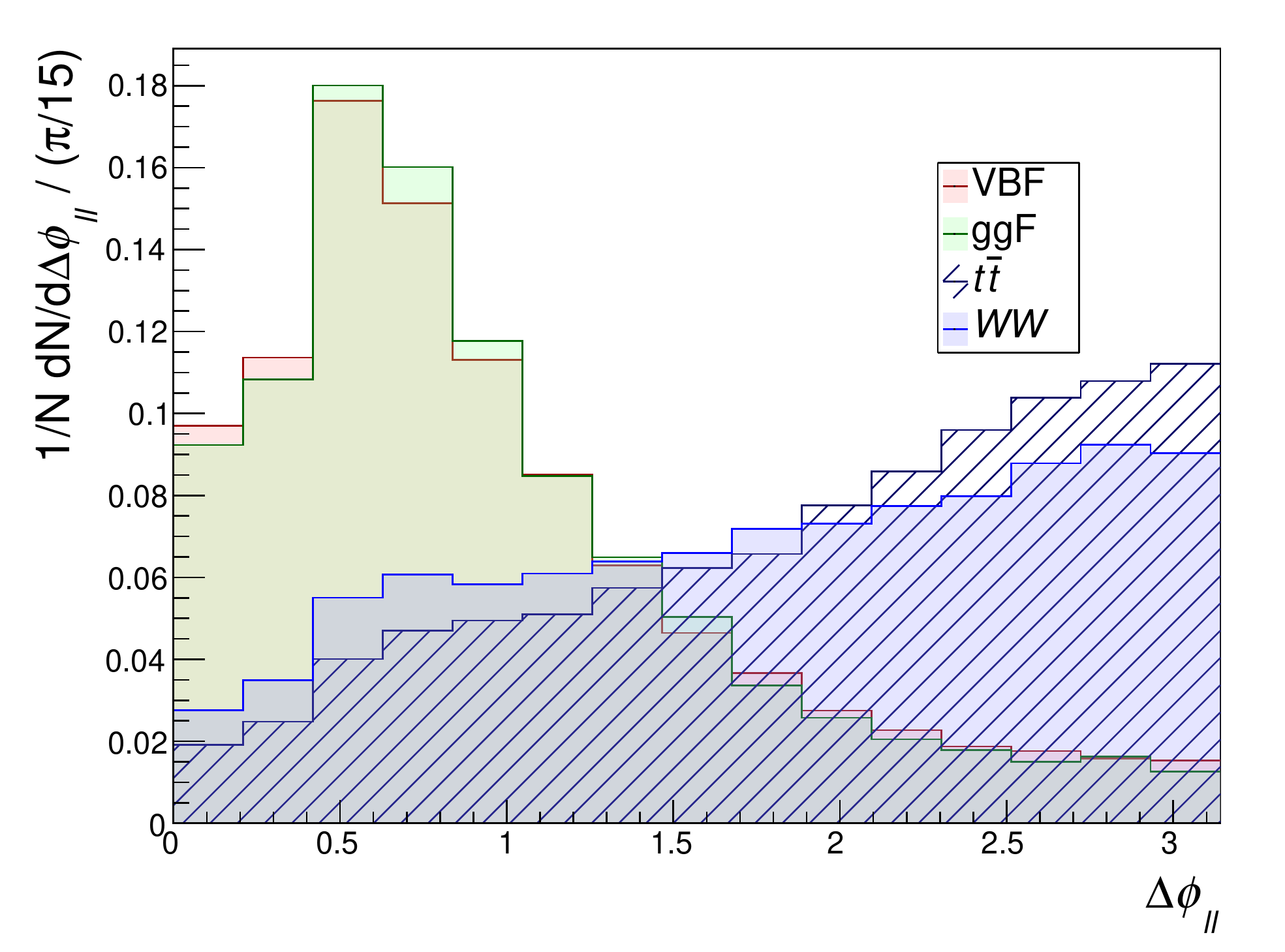}
        \includegraphics[scale=0.35]{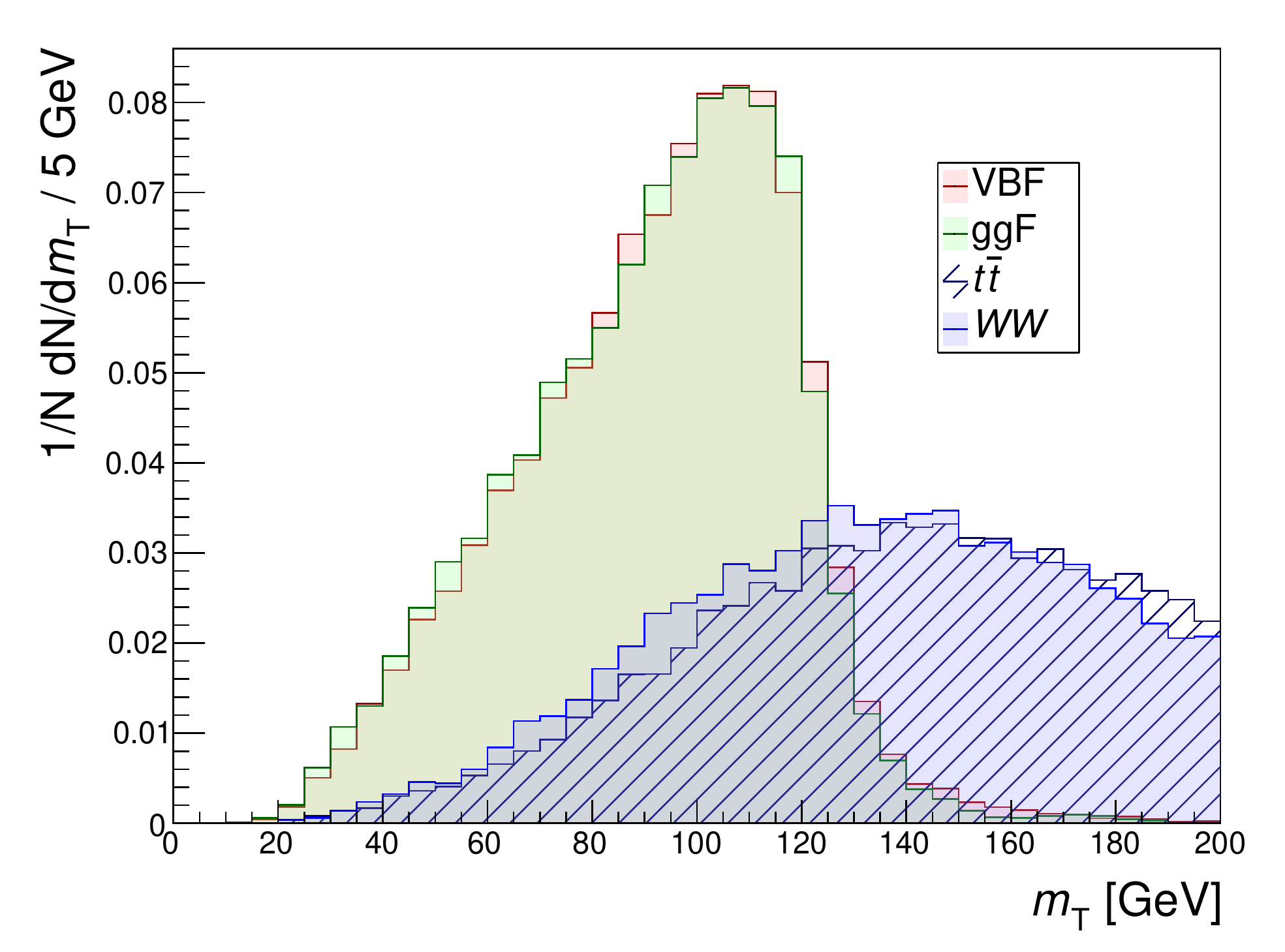}
\caption{Distributions of various variables used in the standard BDT 
in the $H\to WW^*$ channel for the VBF signal, ggF, and the 
SM backgrounds.}
	\label{HWW_dtb}
\end{figure}

\paragraph*{\bf{ Standard BDT}} 
Following the current procedures of the ATLAS analysis, 
the signal sample of VBF and the background samples of 
simulated ggF, simulated $t\bar{t}$, and simulated $WW$ events
are used to train the BDT. We call this one the standard BDT, with
which we shall compare.
The following 8 variables are fed into the BDT:
\begin{enumerate}
\item $m_{jj} $: invariant mass of two leading jets;
\item $\Delta\eta_{jj} \equiv \left|\eta_{j_1}-\eta_{j_2}\right|$;
\item $p_{\mathrm{T}}^{\mathrm{sum}} \equiv p_{\mathrm{T}}^{\ell\ell}+p_{\mathrm{T}}^{\mathrm{miss}}+\sum p_{\mathrm{T}}^{j}$;
\item $\sum m_{\ell_j} \equiv m_{\ell_1,j_1}+m_{\ell_1,j_2}+m_{\ell_2,j_1}+m_{\ell_2,j_2}$;
\item $\sum C_\ell \equiv \sum\left|\eta_\ell-\frac{\sum\eta_{jj}}{2}\right|/\frac{\Delta\eta_{jj}}{2}$;
\item $m_{\ell\ell}$;
\item $\Delta\phi_{\ell\ell}$ ;

\item transverse mass: $m_{\mathrm{T}} \equiv
  \sqrt{(E^{\ell\ell}_\mathrm{T}+p^{\nu\nu}_{\mathrm{T}})^2-\left|\mathbf{p}^{\ell\ell}_{\mathrm{T}}+\mathbf{p}^{\nu\nu}_{\mathrm{T}}\right|^2}$,
  where
  $E^{\ell\ell}_\mathrm{T}=\sqrt{(p^{\nu\nu})^2+(m_{\nu\nu})^2}$,
  $\mathbf{p}^{\nu\nu}(\mathbf{p}^{\ell\ell})$ is the vector sum of
  the neutrino (lepton) transverse momenta, and
  $p^{\nu\nu}(p^{\ell\ell})$is its modulus.

\end{enumerate}
The distributions of these variables for signal and backgrounds
are shown in Fig. \ref{HWW_dtb}, 
in which we can clearly see the capability of each of the variables
in discriminating between the signal and backgrounds.

\begin{figure}[th!]
        \includegraphics[scale=0.4]{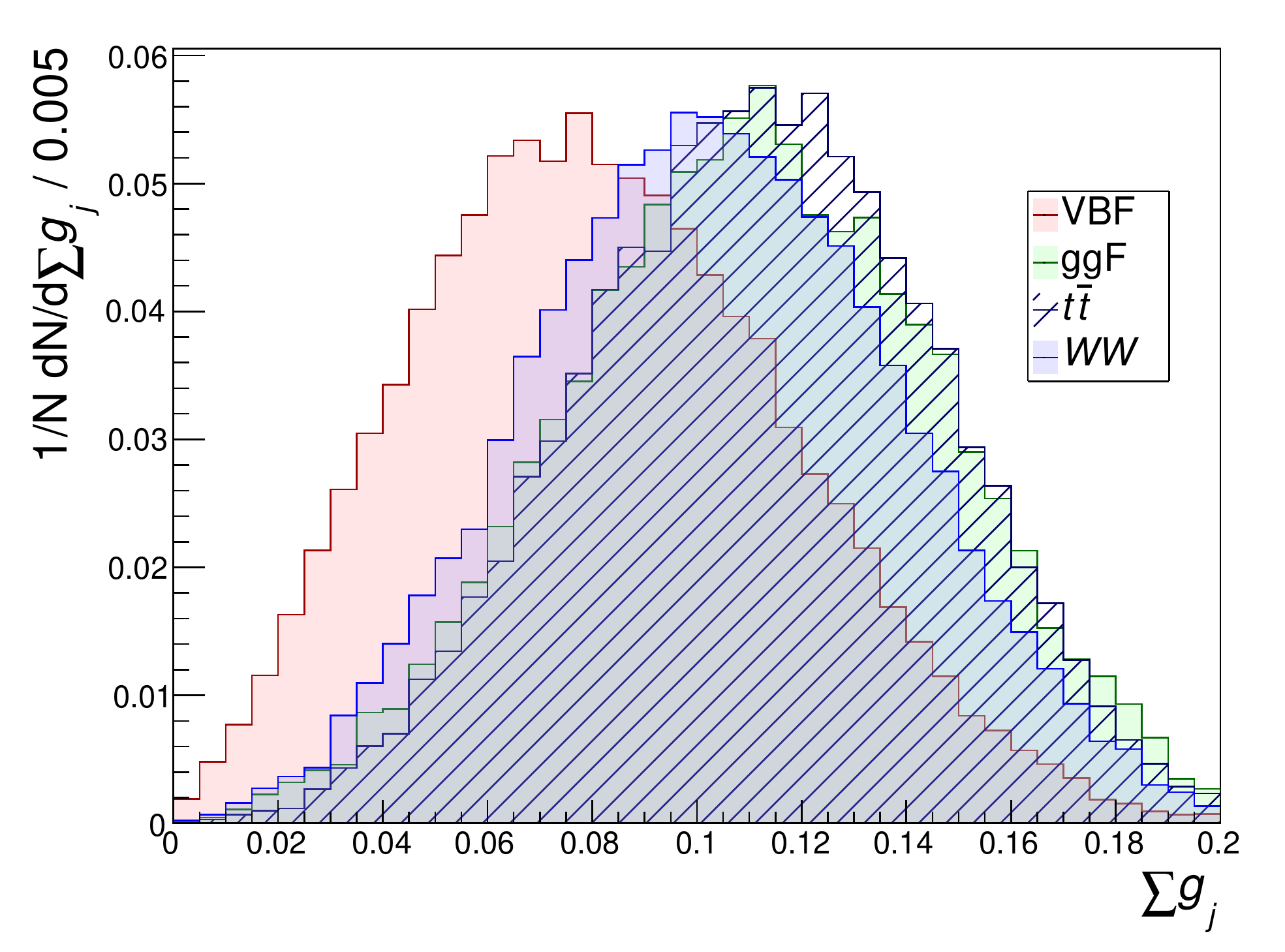}
        \includegraphics[scale=0.4]{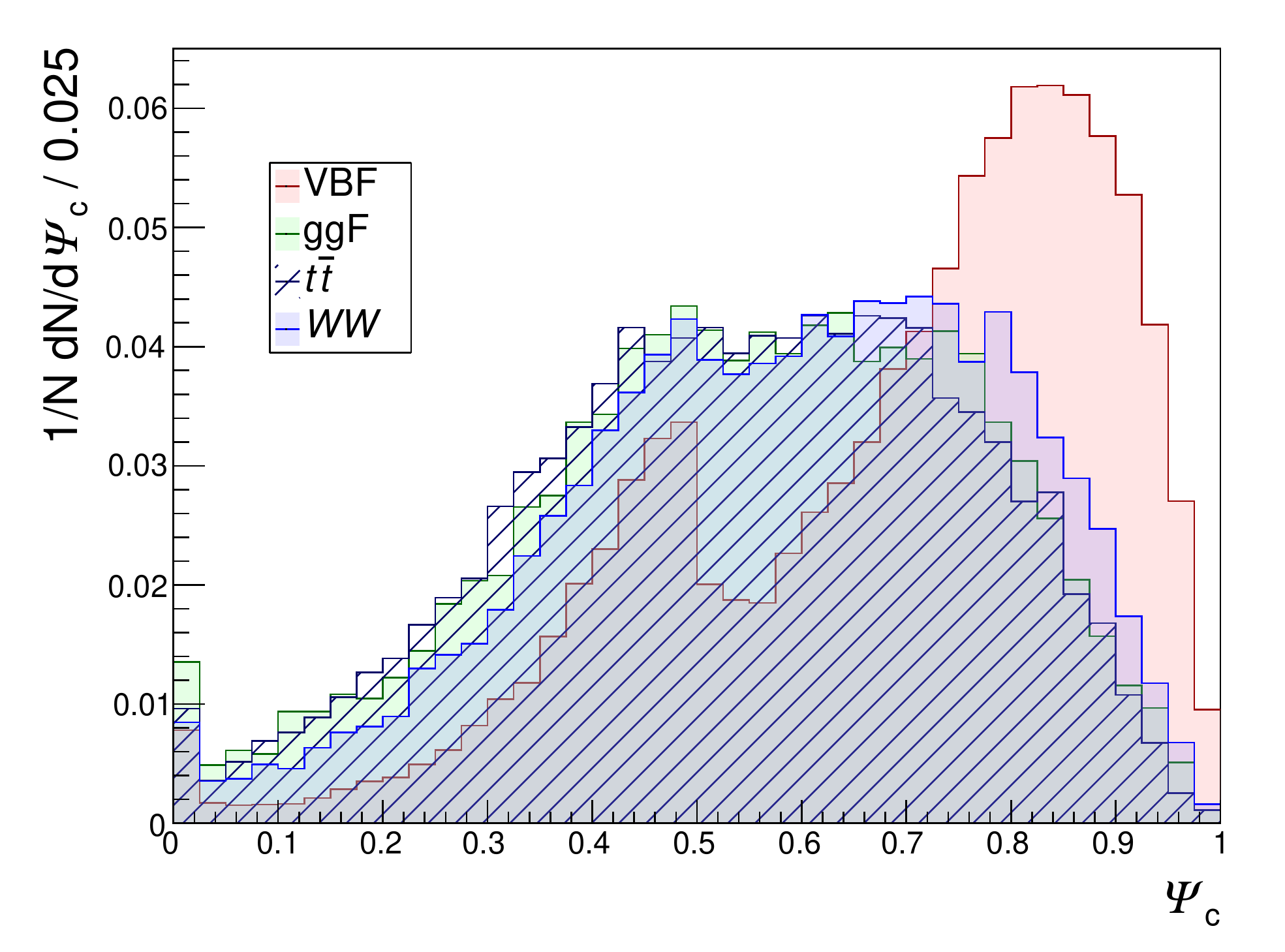}
        \includegraphics[scale=0.4]{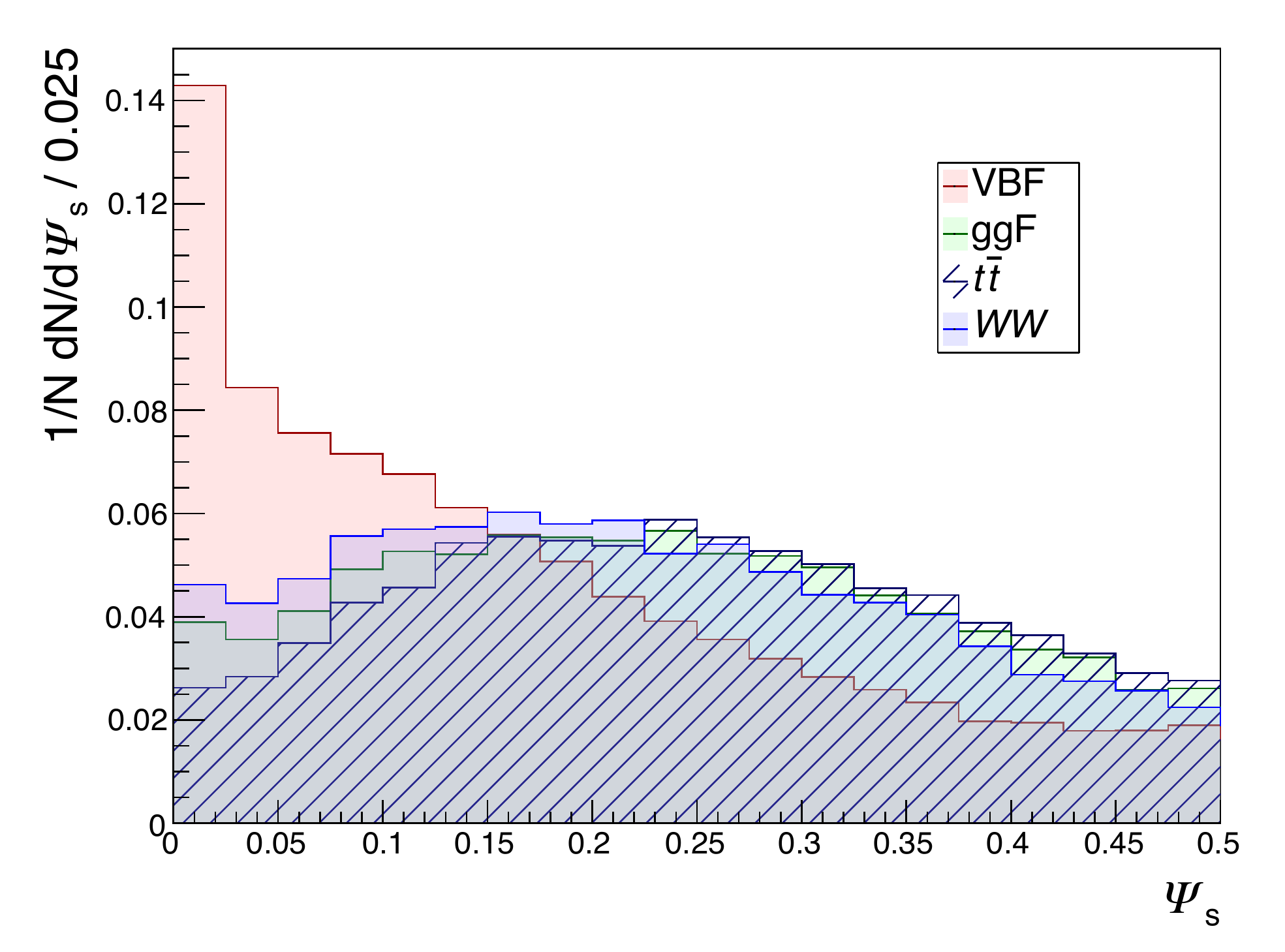}
        \caption{Distributions of the 3 jet-shape variables used 
in the $H\to WW^*$ channel.}
	\label{HWW_dtb_jet_sbstr}
\end{figure}

\paragraph*{{\bf 11-variable BDT}} 
The signal and background training samples are the same as the 
standard BDT. In addition to the 8 variables in standard BDT, 3 more
jet-shape variables \cite{jetshape} 
are employed in this 11-variable BDT analysis:
\begin{enumerate}
 \item girth summed over two leading jets: $\sum g_{j} \equiv
   \underset{j, i \in j}{\sum} \frac{p^{j}_{\mathrm{T}, i}
     r^j_i}{p^{j}_{\mathrm{T}}}$
 \item the central integrated jet shape: $\Psi_{\mathrm{c}} \equiv
   \frac{1}{N}\underset{j=1}{\overset{2}{\sum}} \underset{i\in
     j}{\overset{N}{\sum}}
   \frac{p^j_{\mathrm{T},i}\left(0<r^j_i<0.1\right)}{p^j_{\mathrm{T}}}$
 \item the side integrated jet shape: $\Psi_{\mathrm{s}} \equiv
   \frac{1}{N}\underset{j=1}{\overset{2}{\sum}} \underset{i\in
     j}{\overset{N}{\sum}}
   \frac{p^j_{\mathrm{T},i}\left(0.1<r^j_i<0.2\right)}{p^j_{\mathrm{T}}}$
 \end{enumerate}
 The distributions of these jet-shape variables for the signal and
backgrounds are shown in Fig. \ref{HWW_dtb_jet_sbstr}.
 
 \paragraph*{{\bf 7-variable BDT}} 
Analyzing the distributions shown in Figs. \ref{HWW_dtb} and
Fig. \ref{HWW_dtb_jet_sbstr}, we find that 7 of the variables are 
sufficient in distinguishing between the VBF events and the 
others: $m_{jj}$, $\Delta \eta_{jj}$, $p_{\mathrm{T}}^{\mathrm{sum}}$, $\sum m_{lj}$,
$\sum g_{j}$, $\Psi_{\mathrm{c}}$, and $\Psi_{\mathrm{s}}$. 
The choice of these 7 variables out of the 11 variables is based 
on the ranking output.
Thus, in this
method only these 7 variables are used in discriminating VBF from 
the ggF and backgrounds. The signal and background training
samples are the same as the standard BDT.
		
\paragraph*{{\bf 2-step BDT}} 
This is the new approach that we adopt in this study. We separate the 
training of the BDT in two steps, in which the BDT is trained for VBF
against the SM backgrounds and against the ggF, respectively.
\begin{itemize}
\item 
The first step: the VBF signal sample is trained against 
 the SM background samples of $t\bar{t}$ and $WW$ events. 
 In this step, the variables used are the same as the standard BDT.

\item The second step: after imposing the selection cuts obtained in
the first-step-BDT output $O_{\mathrm{BDT}}^1$, the event samples 
will further undergo the second-step BDT, in which  
the VBF signal sample is trained against the ggF sample only. 
In this step, the variables used are the same as 7-Var BDT.

\end{itemize}

\subsection{$H\to\gamma\gamma$}

Similar to the procedures in $H \to W W^*$, the events samples for
the VBF $H\to\gamma\gamma$ signal, ggF, and the SM background 
have to pass the preselection cuts, which were given in the current 
ATLAS analysis for the SM $H \to \gamma\gamma$ in the VBF
enriched category. The requirements are described as follows:
\begin{enumerate}
\item $N_j \geq 2$;
\item $p_{\mathrm{T}}^{j} > 25 \, \mathrm{GeV}\, \left(\left|\eta^{j}\right|<2.4\right)$ and $p_{\mathrm{T}}^{j} > 30 \, \mathrm{GeV}\, \left(2.4<\left|\eta^{j}\right|<4.4\right)$;
\item  $\Delta \eta_{jj}>2$;
\item $105\leq m_{\gamma\gamma}\leq 160 \,\mathrm{GeV}$;
\item $ p_T^{j1}\geq 0.35 m_{\gamma\gamma}$ and $ p_T^{j2}\geq 0.25 m_{\gamma\gamma}$;
\item $\left|\eta^*\right|<5$, where $\left|\eta^*\right| \equiv \left|\eta_{\gamma\gamma}-\left(\eta_{j1}+\eta_{j2}\right)\right|/2.$ \\
\end{enumerate}

\begin{table}[th!]
     \begin{tabular}{l@{\qquad}l@{\quad}l@{\quad}l@{\quad}l}
    \hline\hline
        Objective & Standard BDT & 9-Var BDT & 5-Var BDT & 2-step BDT  \\
    \hline
    \multirow{1}{*}{Preselection } & \multicolumn{4}{c}{$N_j \geq 2$,}\\
                                                   & \multicolumn{4}{c}{$p_{\mathrm{T}}^{j} > 25 \, \mathrm{GeV}\, \left(\left|\eta^{j}\right|<2.4\right)$ \& $p_{\mathrm{T}}^{j} > 30 \, \mathrm{GeV}\, \left(2.4<\left|\eta^{j}\right|<4.4\right)$,}\\
                                                   & \multicolumn{4}{c}{$\Delta \eta_{jj}>2$,}\\
                                                    & \multicolumn{4}{c}{$105\leq m_{\gamma\gamma}\leq 160 \,\mathrm{GeV}$,}\\
                                                   & \multicolumn{4}{c}{$ p_T^{j1}\geq 0.35 m_{\gamma\gamma}$ and $ p_T^{j2}\geq 0.25 m_{\gamma\gamma}$}\\
                                                   & \multicolumn{4}{c}{$\left|\eta^*\right|<5$}\\
                                                    \hline
     $1^{\mathrm{st}}$ step &&&&\\
    \quad Signal sample & VBF & VBF & VBF & VBF\\
    \quad Bkg. sample & ggF \& $\gamma\gamma+jj$&ggF \& $\gamma\gamma+jj$ &ggF \& $\gamma\gamma+jj$\,\,\,\,\,\,& $\gamma\gamma+jj$\\
    \multirow{1}{*}{\quad BDT inputs}  & $m_{jj} $, $\Delta\eta_{jj}$, $p_{\mathrm{Tt}}$, & $m_{jj} $, $\Delta\eta_{jj}$, $p_{\mathrm{Tt}}$, & $m_{jj} $, $\Delta\eta_{jj}$, & $m_{jj} $, $\Delta\eta_{jj}$, $p_{\mathrm{Tt}}$,\\
                                  & $\Delta R^{\mathrm{min}}_{\gamma,j}$, $\left|\eta^*\right|$, $\phi^*$ & $\Delta R^{\mathrm{min}}_{\gamma,j}$, $\left|\eta^*\right|$, $\phi^*$, & $\sum g_{j}$, $\Psi_{\mathrm{c}}$, $\Psi_{\mathrm{s}}$, & $\Delta R^{\mathrm{min}}_{\gamma,j}$, $\left|\eta^*\right|$, $\phi^*$,\\
                               & & $\sum g_{j}$, $\Psi_{\mathrm{c}}$, $\Psi_{\mathrm{s}}$ & &\\
                                                                 \hline
    $2^{\mathrm{nd}}$ step &&&&\\
   \quad Signal sample  & \quad - & \quad - & \quad - & VBF \\
\quad Bkg. sample  & \quad - & \quad - & \quad - & ggF\\
     \multirow{1}{*}{\quad BDT inputs}  & \multirow{2}{*}{\quad -} & \multirow{2}{*}{\quad -} & \multirow{2}{*}{\quad -} & $m_{jj}$, $\Delta \eta_{jj}$,  \\
                           & & & & $\sum g_{j}$, $\Psi_{\mathrm{c}}$, $\Psi_{\mathrm{s}}$, \\
   \hline\hline
    \end{tabular}
    \caption{Summary of each analytic method for $H\to \gamma\gamma$  }
    \label{tab:BDT_Haa}
    \end{table}

\begin{figure}[th!]
        \includegraphics[scale=0.4]{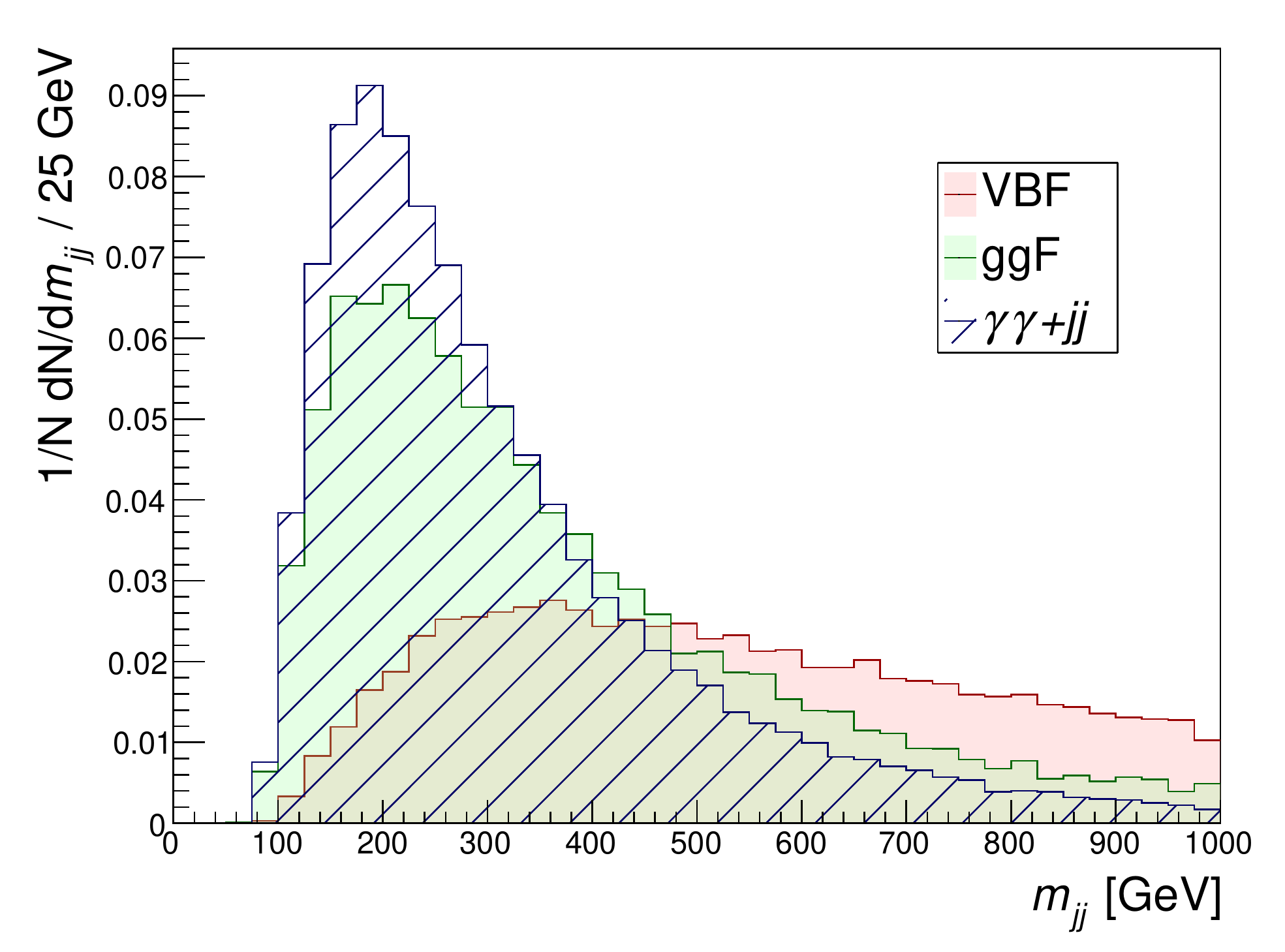}
        \includegraphics[scale=0.4]{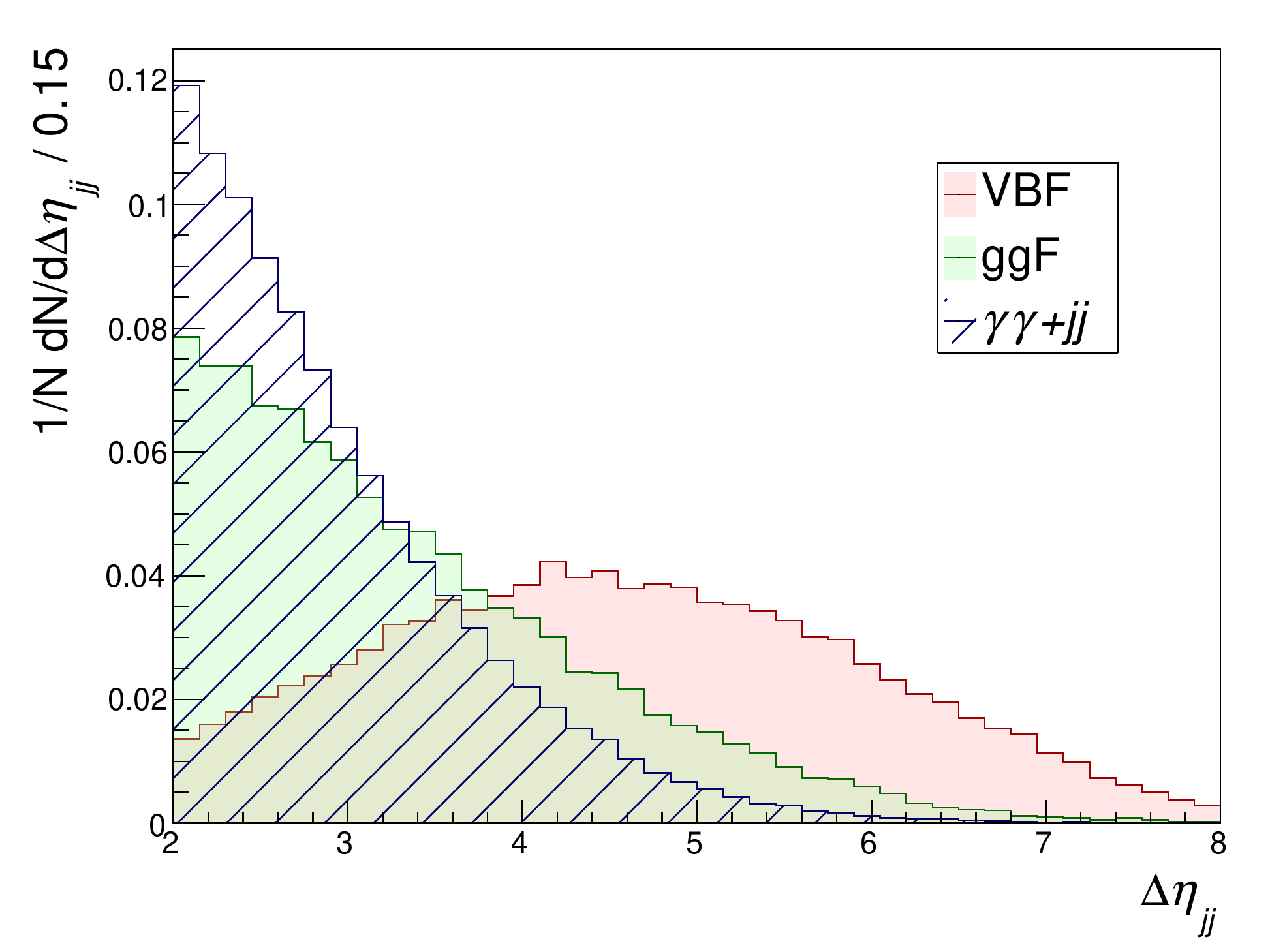}
        \includegraphics[scale=0.4]{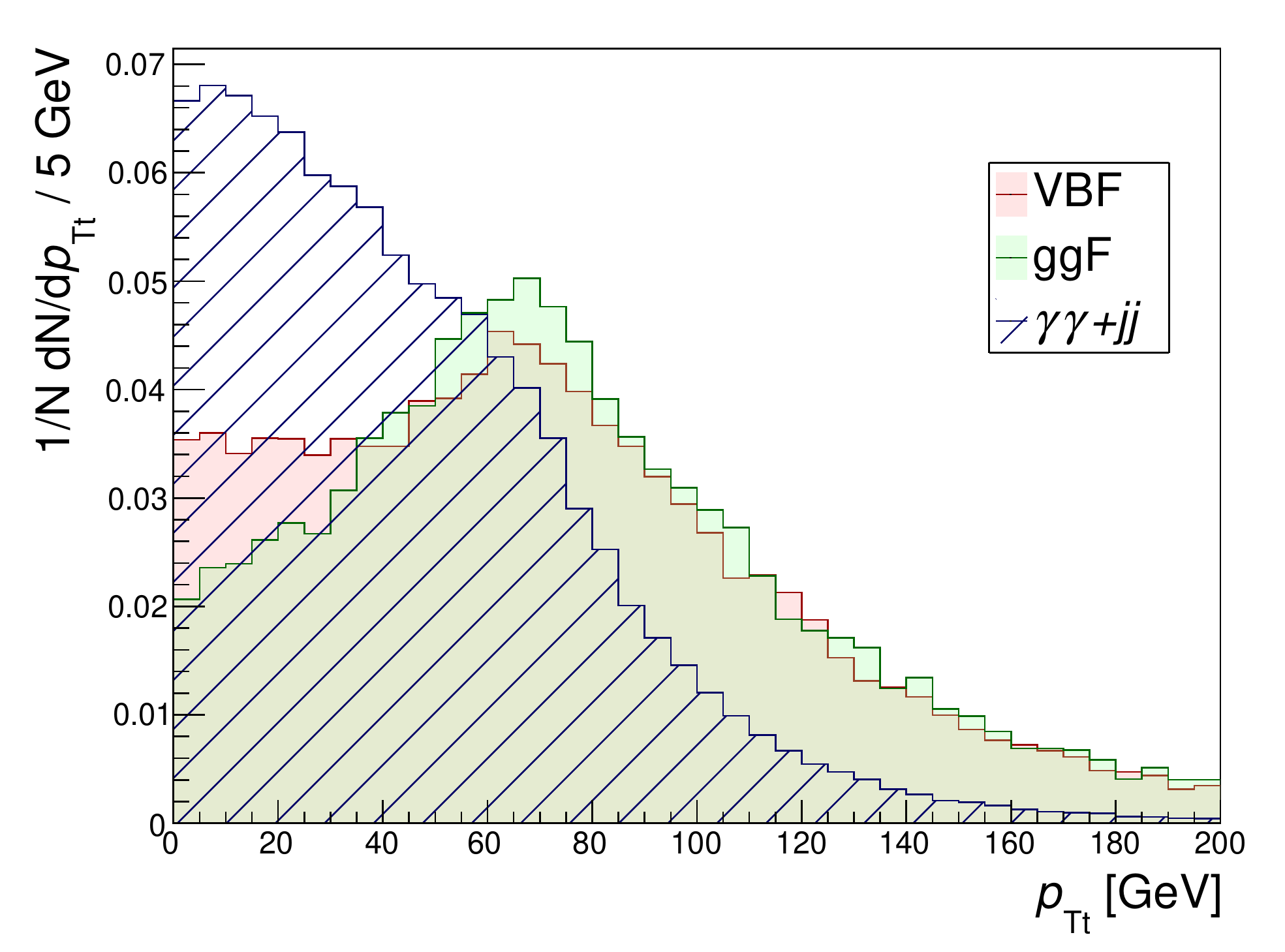}
        \includegraphics[scale=0.4]{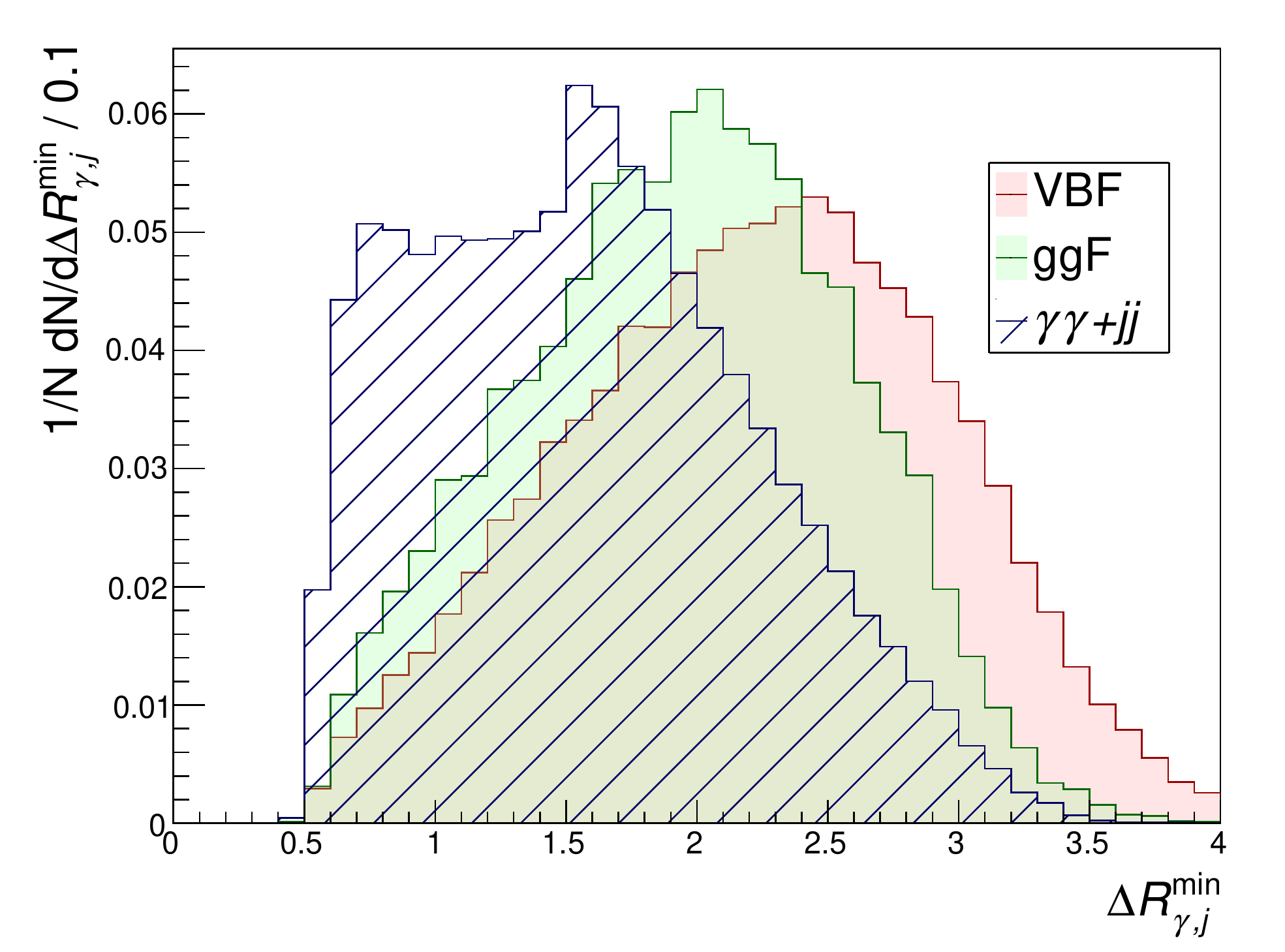}
        \includegraphics[scale=0.4]{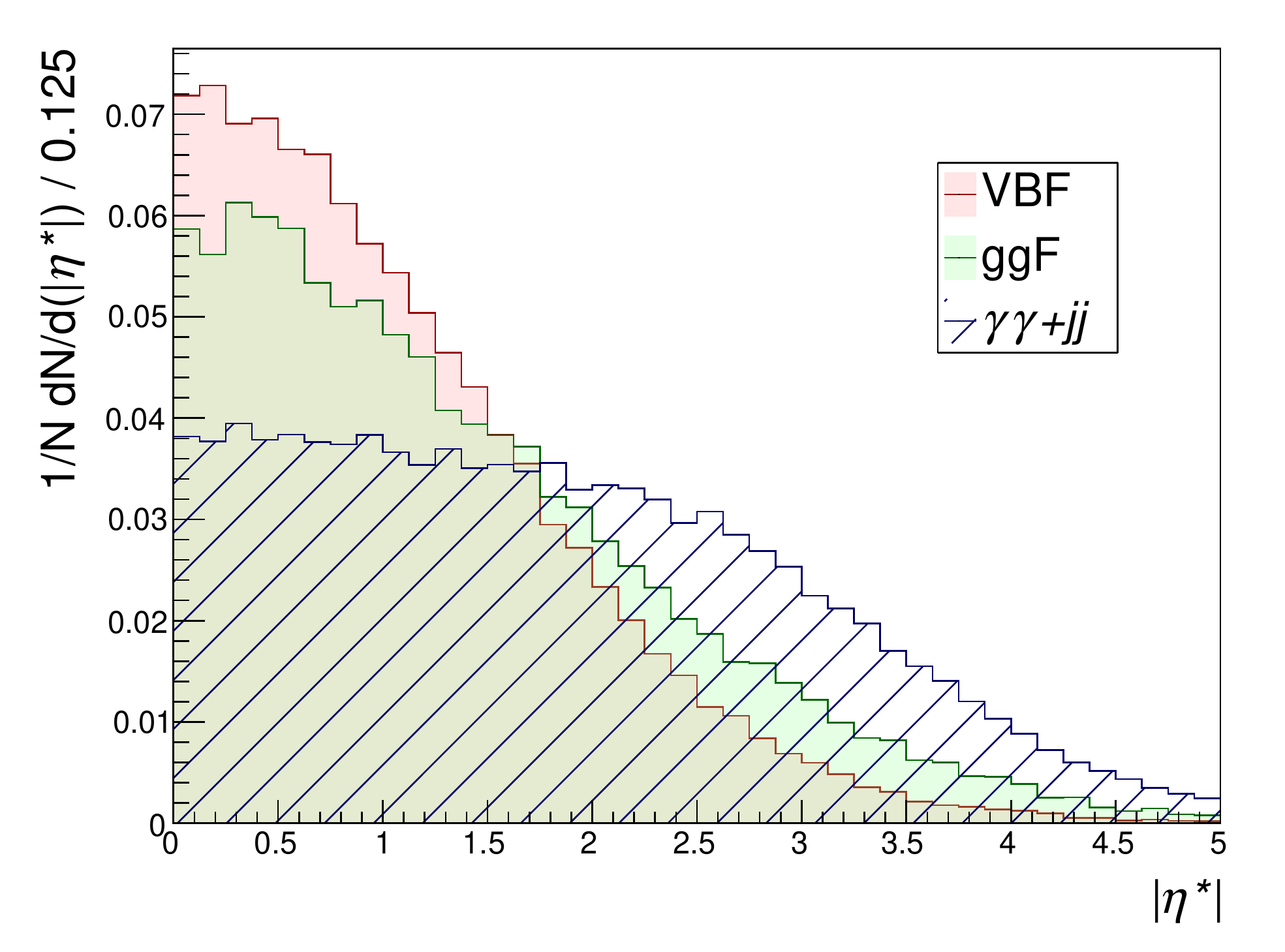}
        \includegraphics[scale=0.4]{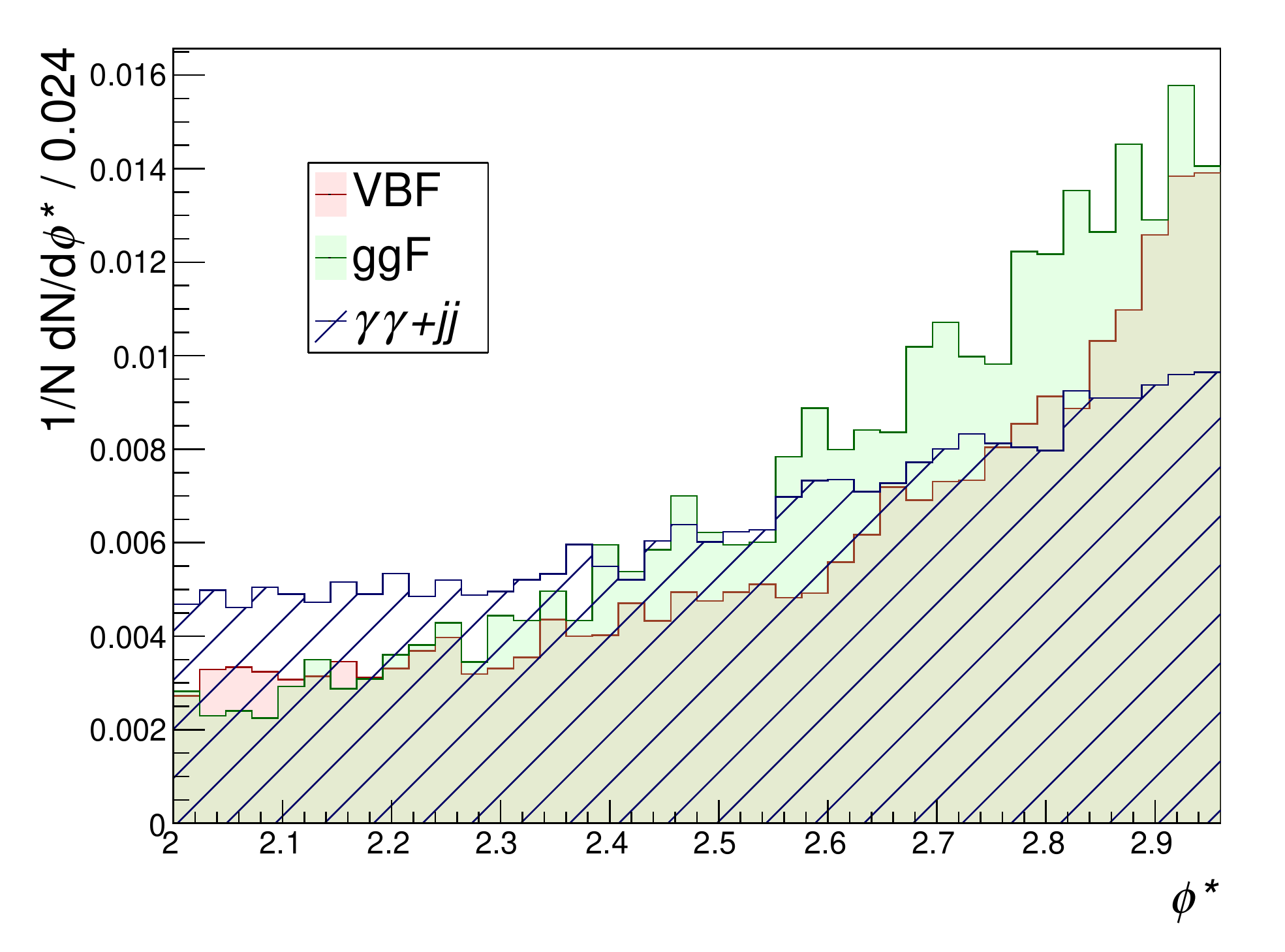}
        \caption{Distributions of the 6 variables used in the channel
$H\to \gamma\gamma$ for the VBF signal, ggF, and the SM background
$\gamma\gamma jj$.}
	\label{Haa_dtb}
\end{figure}

\paragraph*{\bf{ Standard BDT}} 
Following the current procedures in the ATLAS analysis, the signal
sample of VBF and the background samples of ggF events and simulated
$\gamma\gamma+jj$ events are used to train the BDT.
Again, this is the standard BDT.
The following 6 variables are inputs to the BDT:
\begin{enumerate}
\item $m_{jj} $;
\item $\Delta\eta_{jj}$;
\item $p_{\mathrm{Tt}} \equiv
  \left|\left(\mathbf{p}_\mathrm{T}^{\gamma_1}+\mathbf{p}_\mathrm{T}^{\gamma_2}\right)\times\hat{t}\right|$,
  where
  $\hat{t}=\left(\mathbf{p}_\mathrm{T}^{\gamma_1}-\mathbf{p}_\mathrm{T}^{\gamma_2}\right)/\left|\mathbf{p}_\mathrm{T}^{\gamma_1}-\mathbf{p}_\mathrm{T}^{\gamma_2}\right|$;
\item $\Delta R^{\mathrm{min}}_{\gamma,j} \equiv $ the minimum
  separation between the leading/subleading photon and the
  leading/subleading jet;
\item $\left|\eta^*\right|$;
\item $\phi^* \equiv$ the azimuthal angle between the diphoton and the
  dijet system.
\end{enumerate}
The distributions of these variables for the signal and backgrounds
are shown in Fig. \ref{Haa_dtb}.

\begin{figure}[th!]
        \includegraphics[scale=0.4]{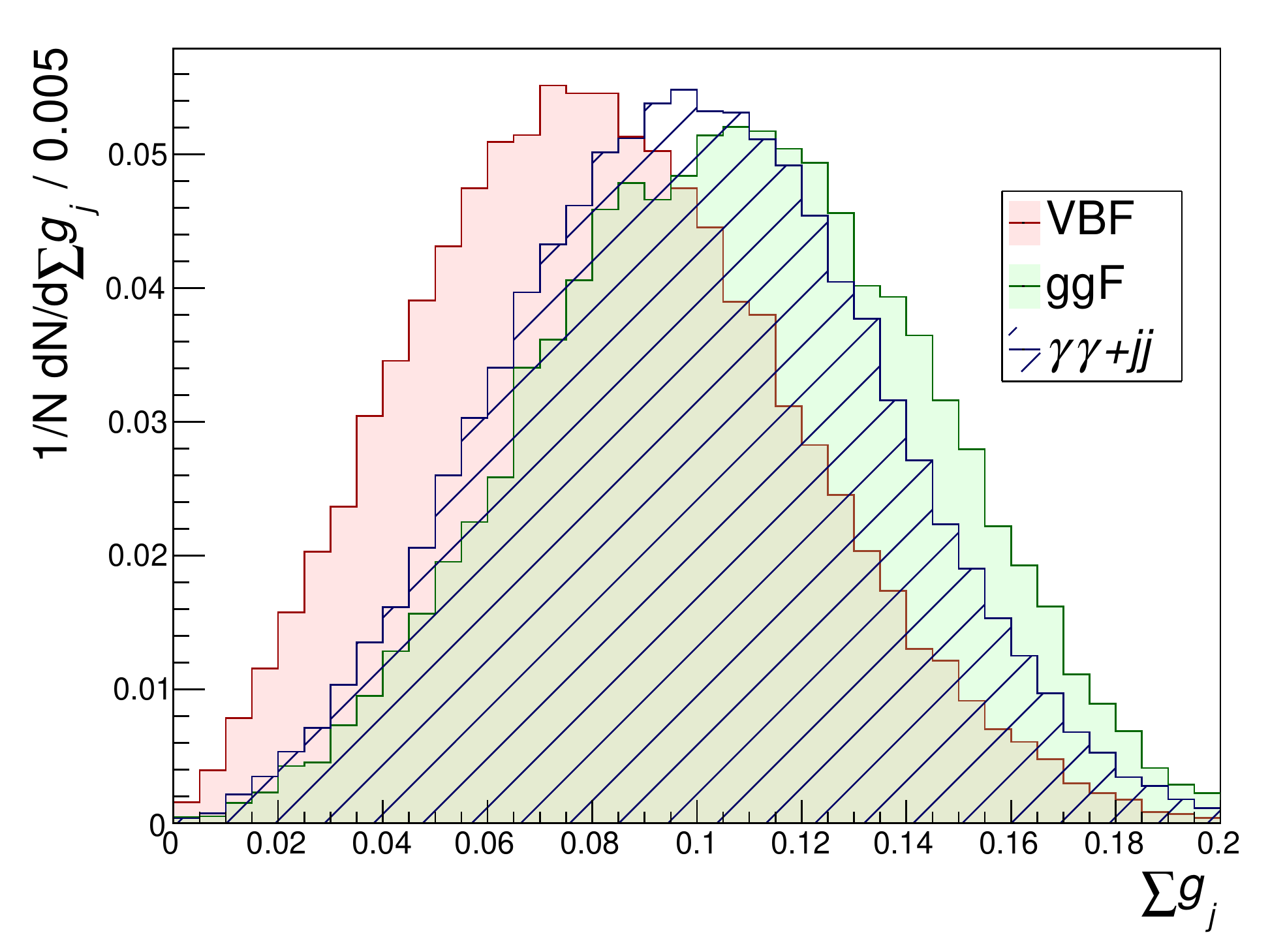}
        \includegraphics[scale=0.4]{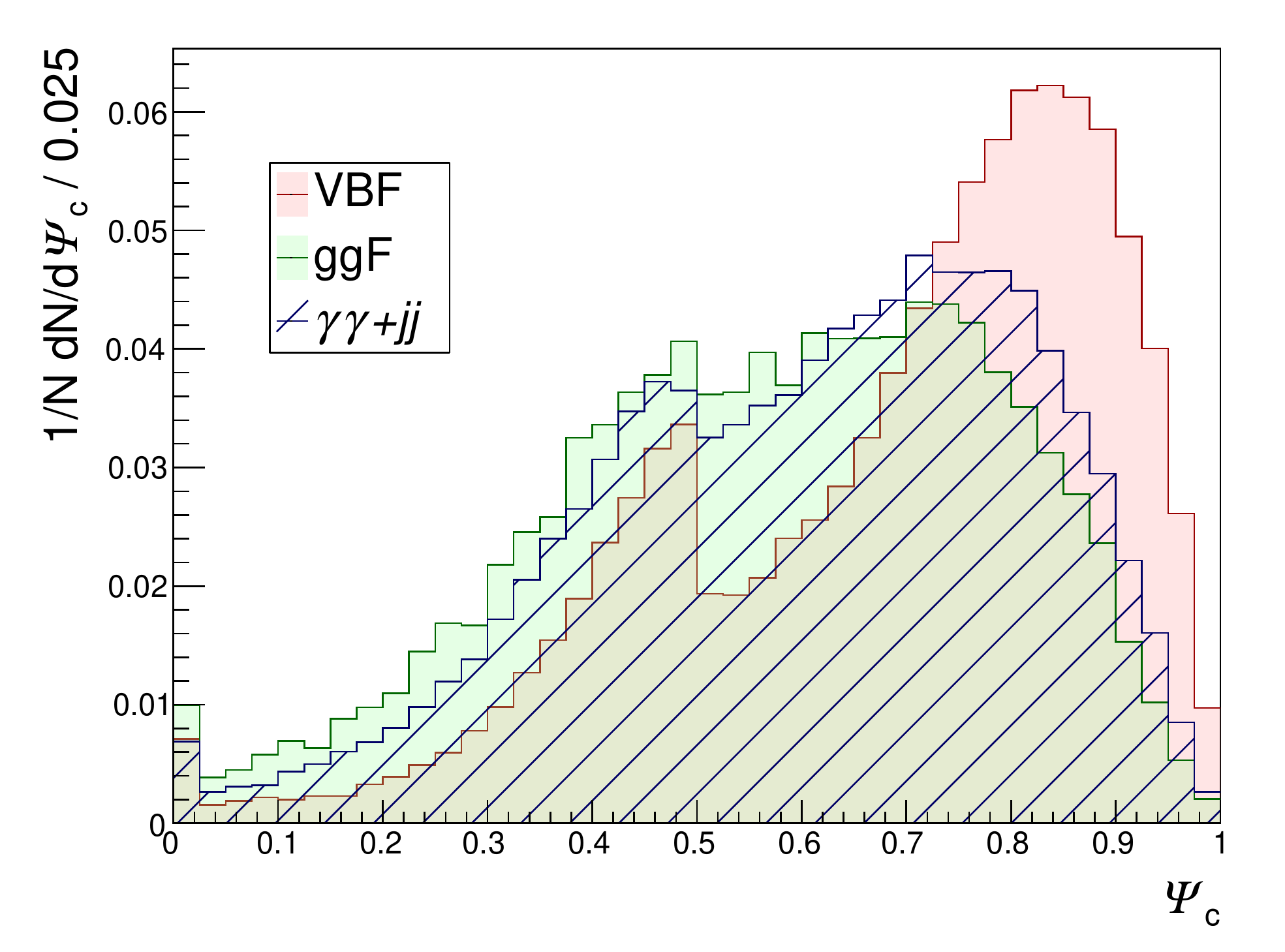}
        \includegraphics[scale=0.4]{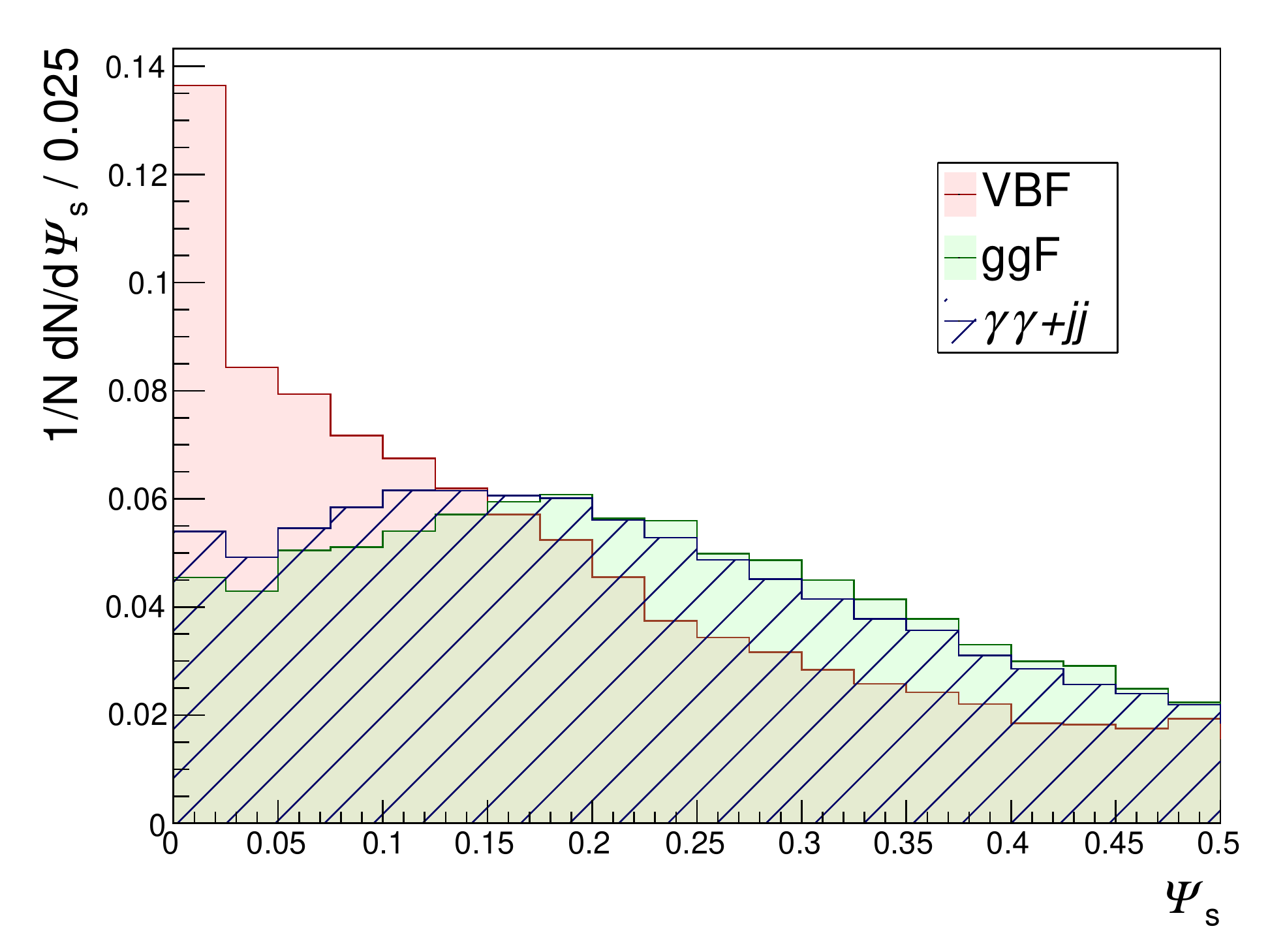}
        \caption{Distributions of the 3 jet-shape variables 
used in the channel $H\to \gamma\gamma$.}
	\label{Haa_dtb_jet_sbstr}
\end{figure}

\paragraph*{{\bf 9-variable BDT}} 
The signal and background training samples are the same as the
standard BDT. In addition to the 6 variables in the standard BDT, 3 more
jet-shape variables are used in this 9-variable BDT: $\sum 
g_{j}$, $\Psi_{c}$, $\Psi_{s}$, whose distributions are
shown in Fig. \ref{Haa_dtb_jet_sbstr}.

\paragraph*{{\bf 5-variable BDT}} 
Analyzing the distributions of the above 9 variables
we find five most powerful variables in
discriminating between VBF and ggF.  They are $m_{jj} $, $\Delta\eta_{jj}$,
$\sum g_{j}$, $\Psi_{\mathrm{c}}$, $\Psi_{\mathrm{s}}$, as shown in
Fig. \ref{Haa_dtb} and Fig. \ref{Haa_dtb_jet_sbstr}.

\paragraph*{{\bf 2-step BDT}} 
Again, this is the new approach that we are adopting in this study.
We separate the training of the BDT in two steps:
\begin{itemize}
\item The first step: the VBF signal sample is trained against 
the background sample of $\gamma\gamma+jj$ events. 
In this step, the variables used are the same as the standard BDT.

\item The second step: 
after imposing the selection cuts obtained in
the first-step-BDT output $O_{\mathrm{BDT}}^1$, the event samples 
will further undergo the second-step BDT, in which
the VBF signal samples is trained against the ggF sample. 
In this step, the variables used are the same as 5-Var BDT.
\end{itemize}

\section{Results}

\subsection{$H\to WW^*\to e\nu\mu\nu$}

\begin{figure}[th!]
        \includegraphics[scale=0.4]{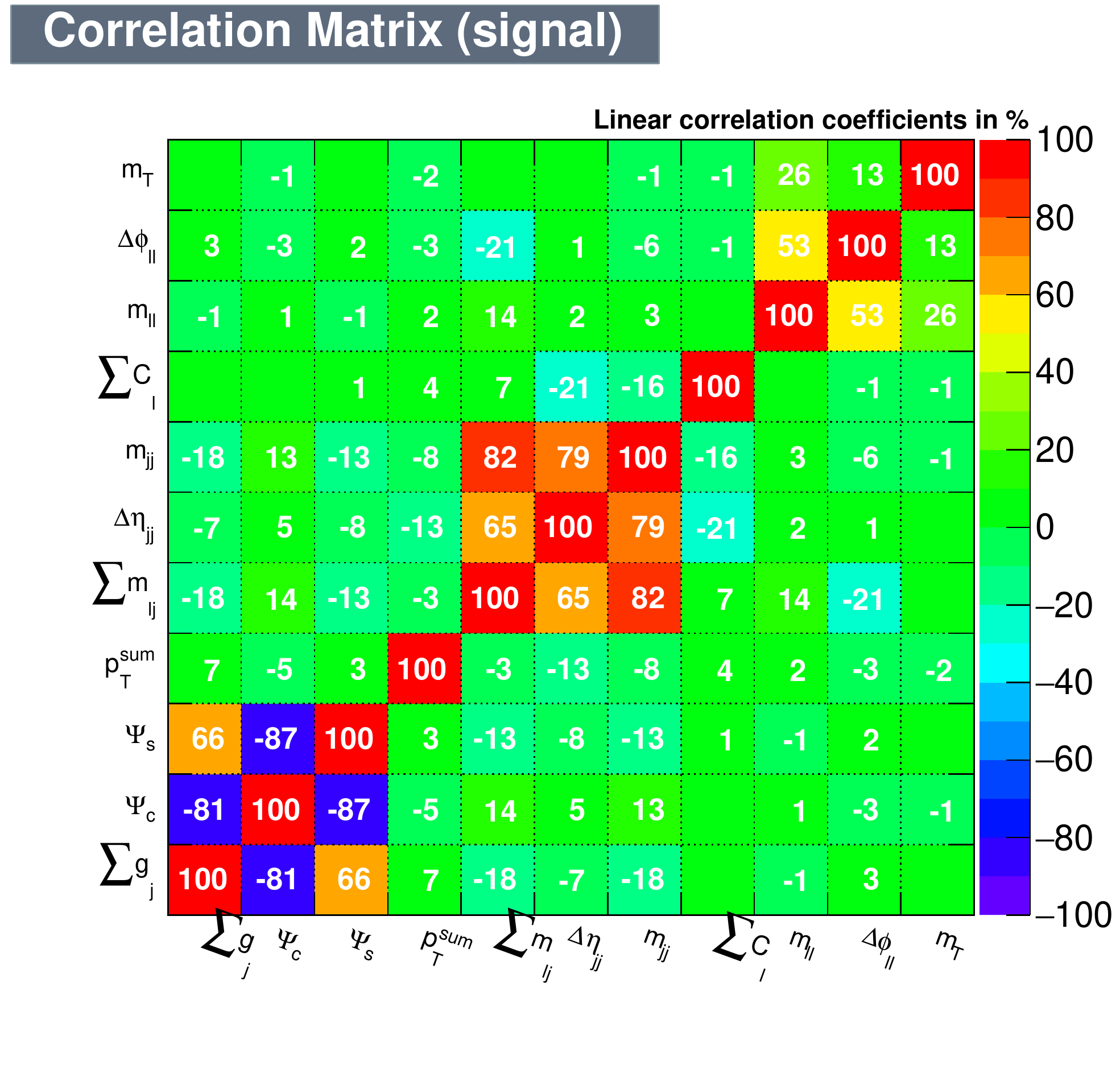}
        \includegraphics[scale=0.4]{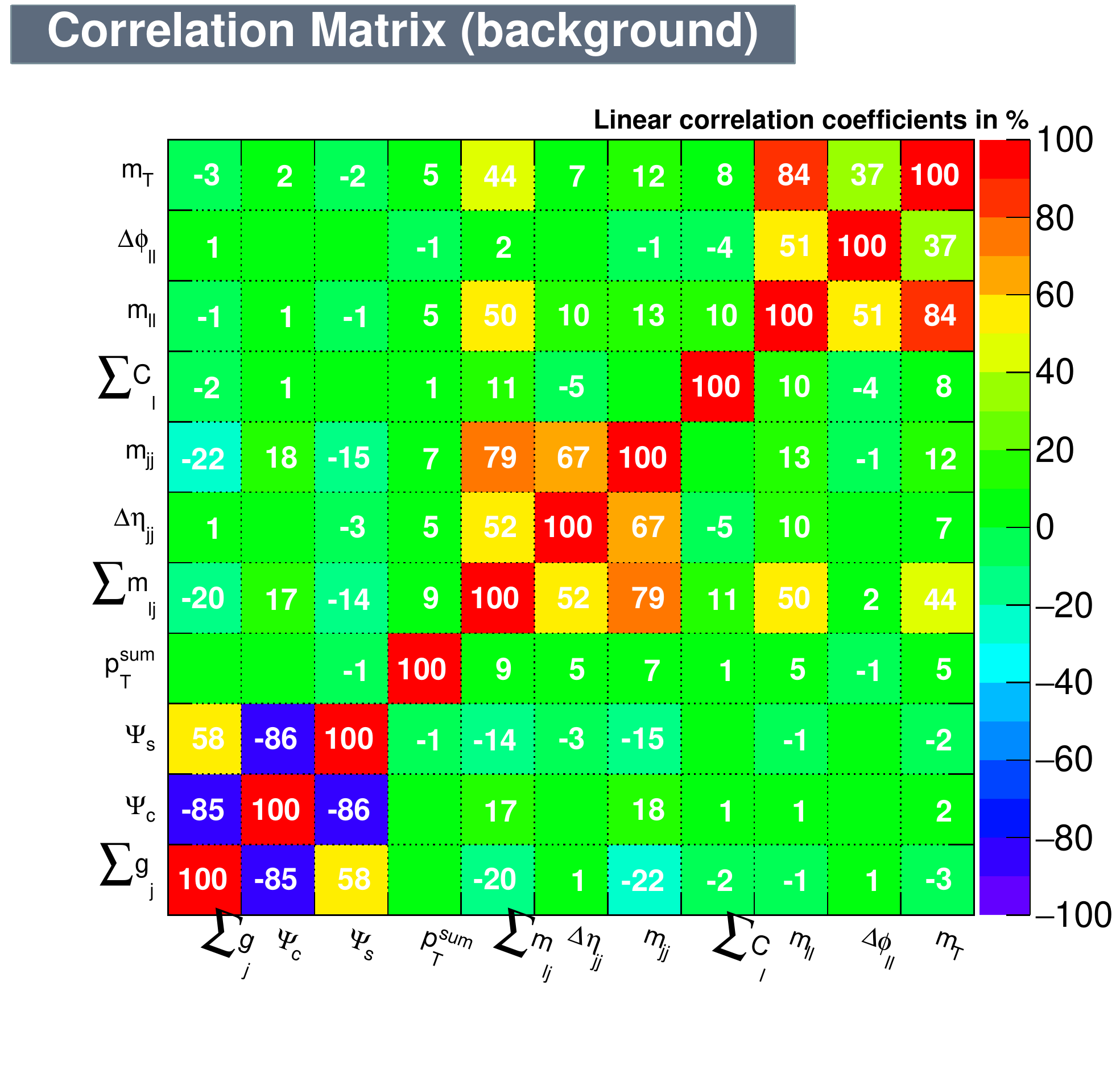}
        \caption{Linear correlations between any two variables 
in the channel $H\to WW^*\to e\nu\mu\nu$ used in the
11-Var BDT. The left panel shows the VBF while the right panel
includes ggH, $t\bar{t}$, and $WW$.}
	\label{HWW_Corr}
\end{figure}

\begin{figure}[th!]
        \includegraphics[scale=0.4]{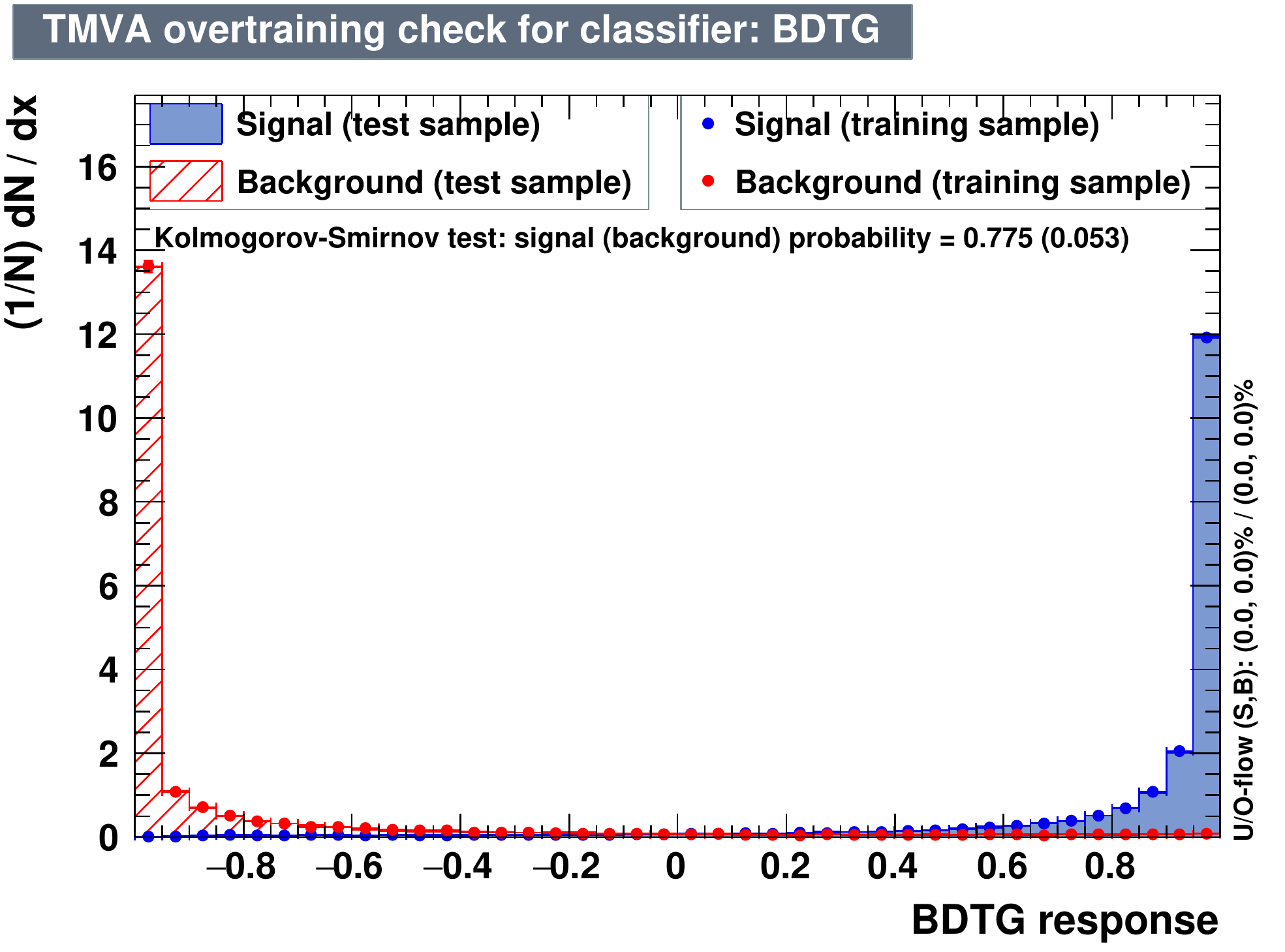}
        \includegraphics[scale=0.4]{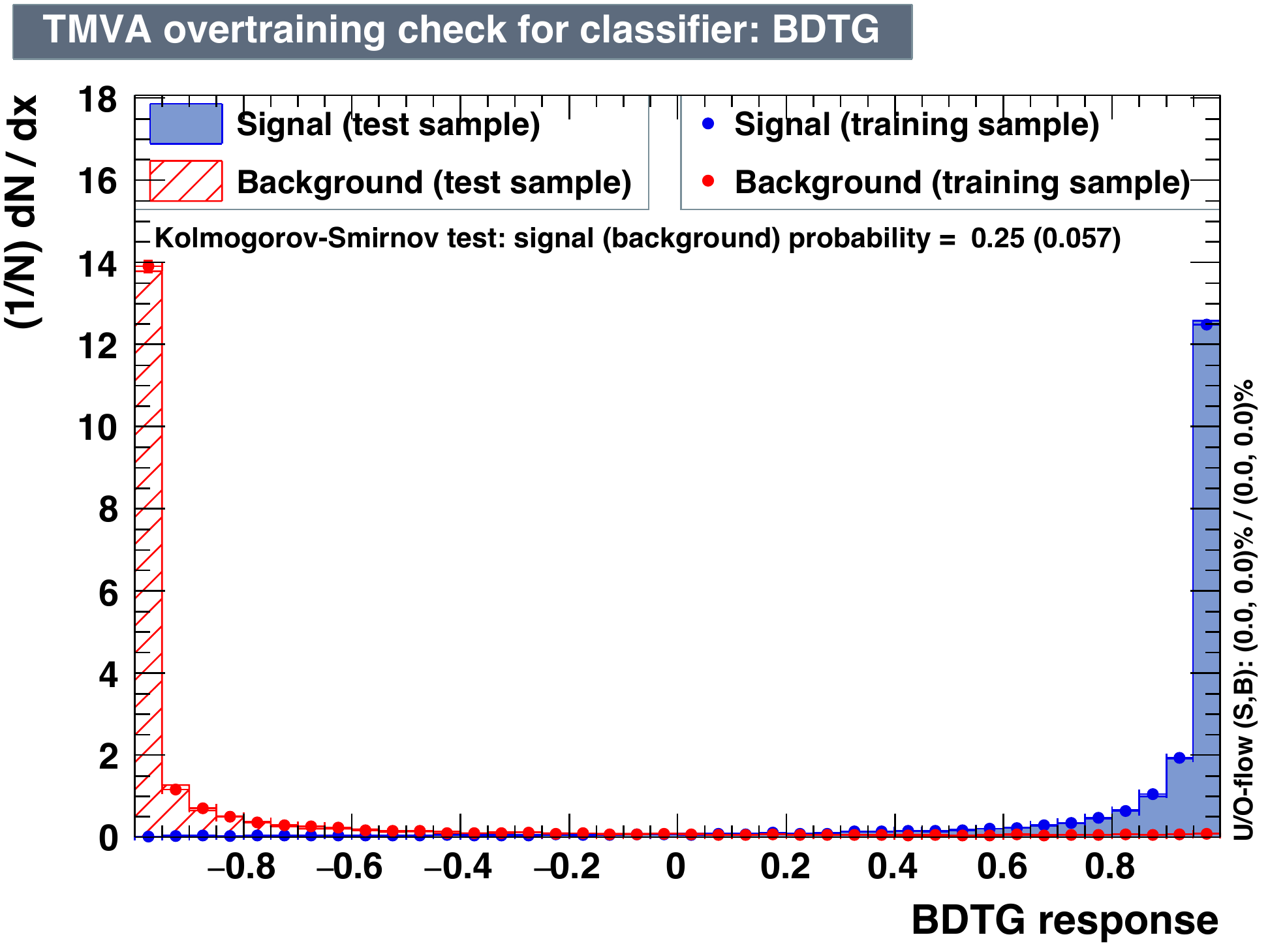}
        \includegraphics[scale=0.4]{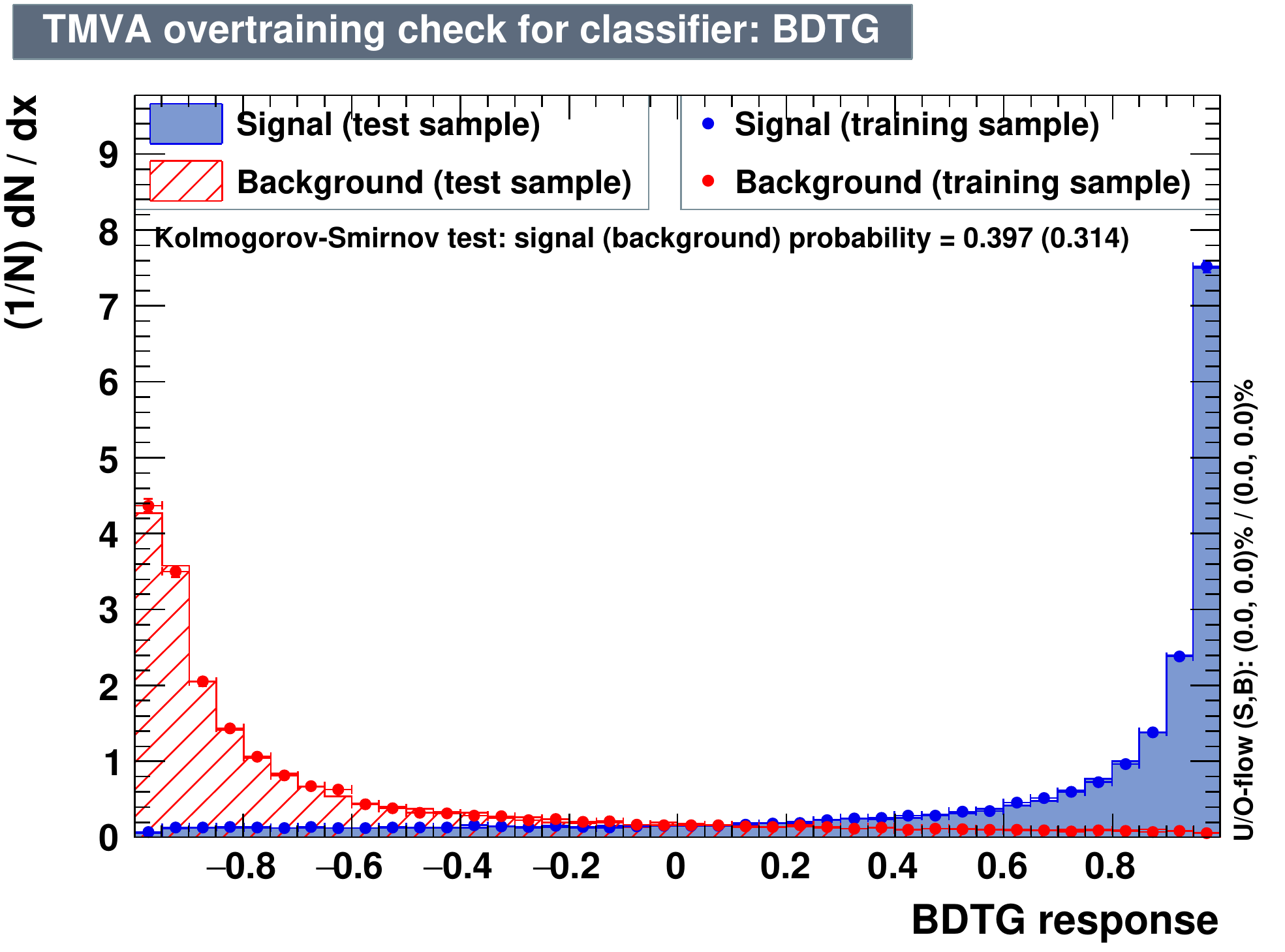}
        \includegraphics[scale=0.4]{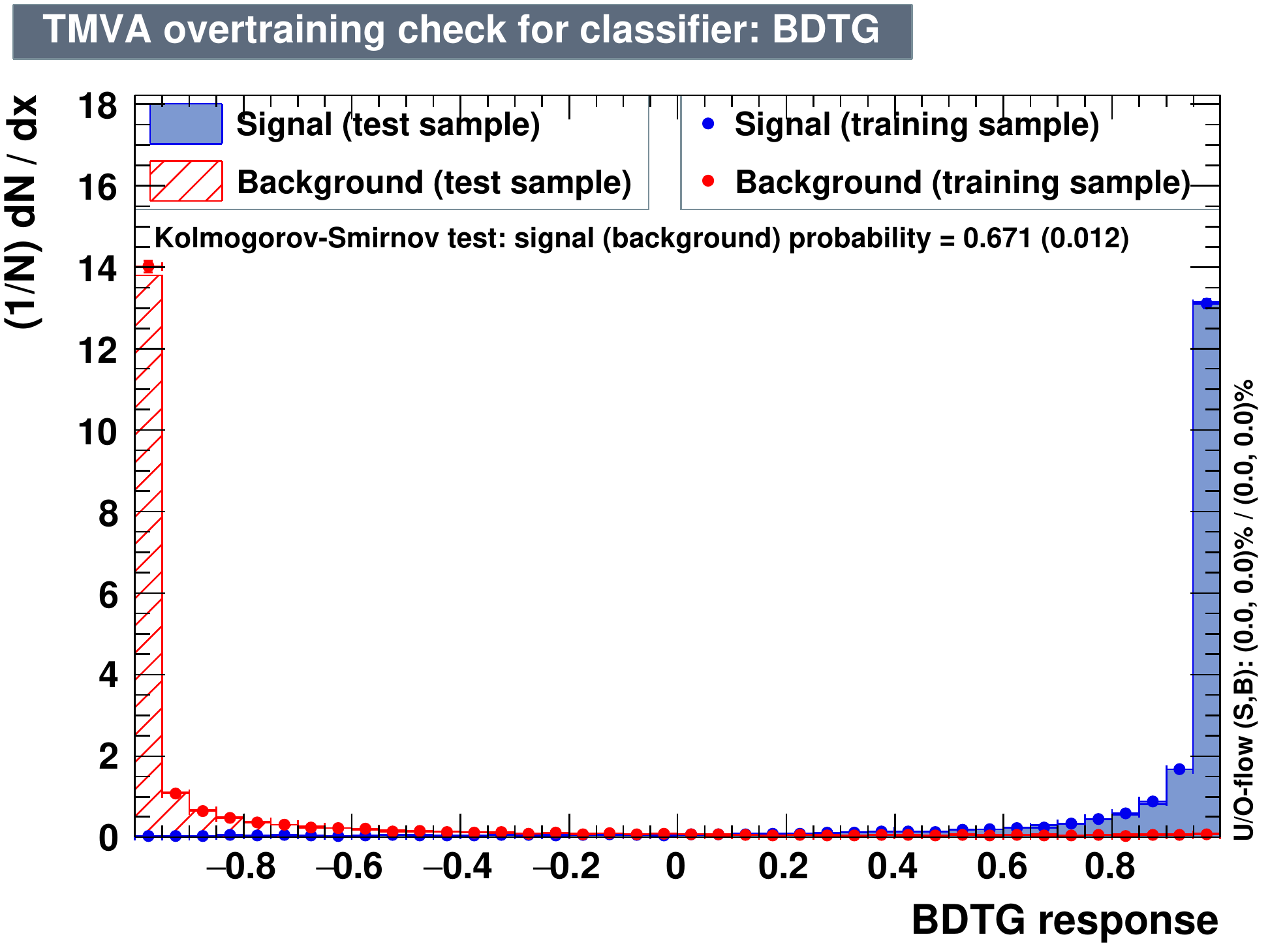}
        \includegraphics[scale=0.4]{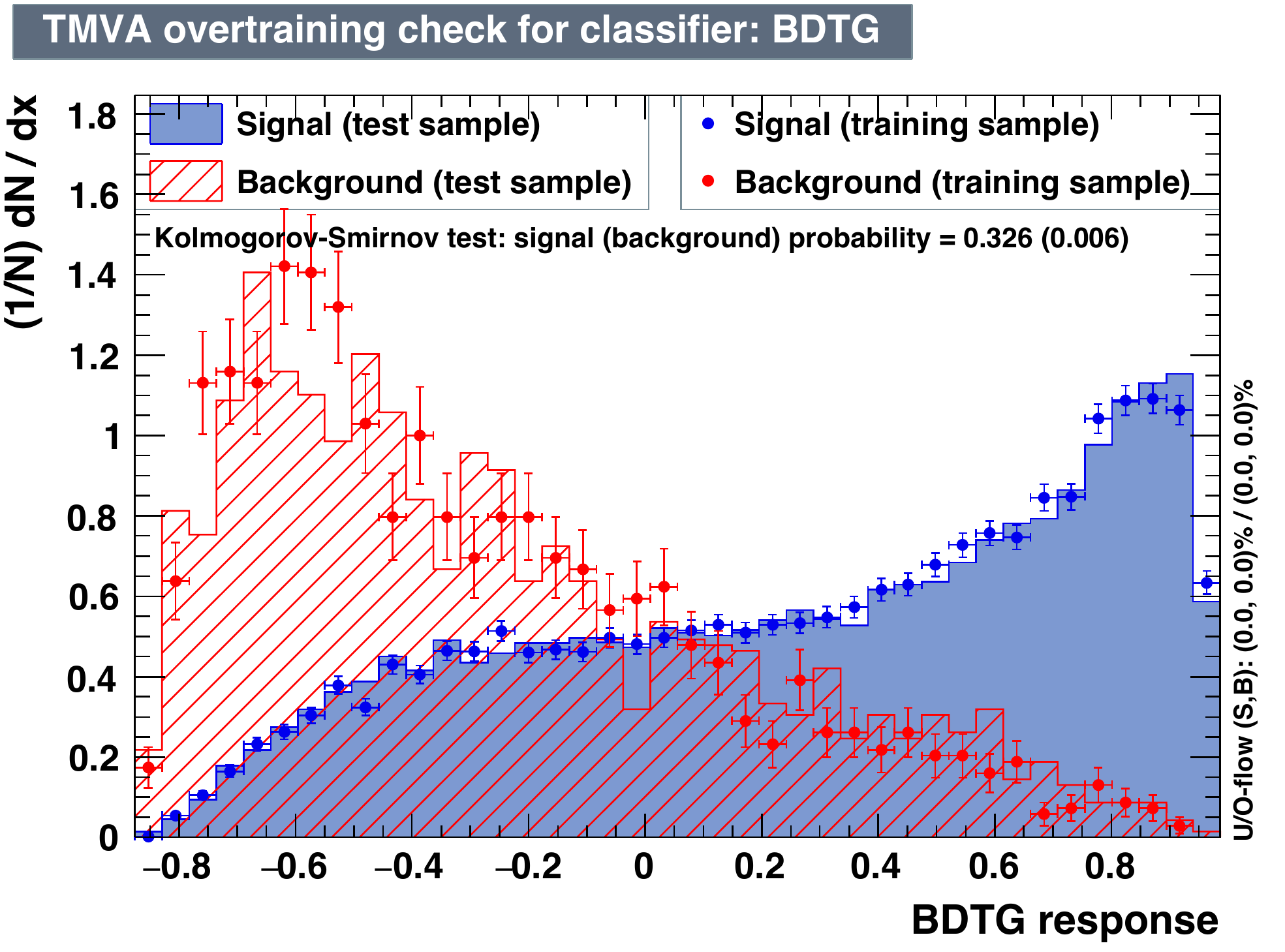}
        \caption{BDT output distributions of training and test samples
          in $H\to WW^*$. (\textit{Top-left}) Method of standard
          BDT. (\textit{Top-right}) Method of 11-Var
          BDT. (\textit{Middle-left}) Method of 7-Var
          BDT. (\textit{Middle-right}) The first step in the 2-step
          BDT. (\textit{Bottom}) The second step in the 2-step BDT. 
Note that half of sample is used as the training sample and the other 
half as the testing sample.}
	\label{HWW_BDT}
\end{figure}

Figure \ref{HWW_Corr} shows the linear correlations between any two
of the variables used in the 11-Var BDT for the channel $H\to WW^*$. 
{}From the figure we can see very strong correlations 
appear among the 3 jet-shape variables, and 
among $\sum m_{\ell j}$, $\Delta \eta_{jj}$, and $m_{jj}$
in both the signal and backgrounds. 
A sizeable correlation also appears between $m_{\ell\ell}$ and 
$\Delta \phi_{\ell\ell}$ in both the signal and
backgrounds.
In addition, in order to avoid overtraining in
BDT analyses, we show the BDT output distributions for both the
training and testing samples  in Fig. \ref{HWW_BDT}.

The results of our analyses for the channel $H\to WW^*$ 
are summarized in Table \ref{HWW_result}. 
The final numbers of the remained VBF events for all methods
are all around $5.1$, 
in order to have direct comparisons among various methods used 
here.
Comparing between the standard BDT and the 11-Var BDT, the latter
which used 3 jet-shape variables, can enhance the VBF purity and 
at the same time reduce the ggF contamination by about $2 \%$. 
When we focus on distinguishing just between the VBF and ggF event
samples, the 7-Var BDT using the most powerful 7
variables is introduced and can further decrease the ggF
contamination by about $1 \%.$ 
However, this method sacrifices 
the discrimination between the VBF sample and the other SM backgrounds,
and thus lowers the VBF purity to only $50.5\%$.

\begin{figure}[th!]
        \includegraphics[scale=0.4]{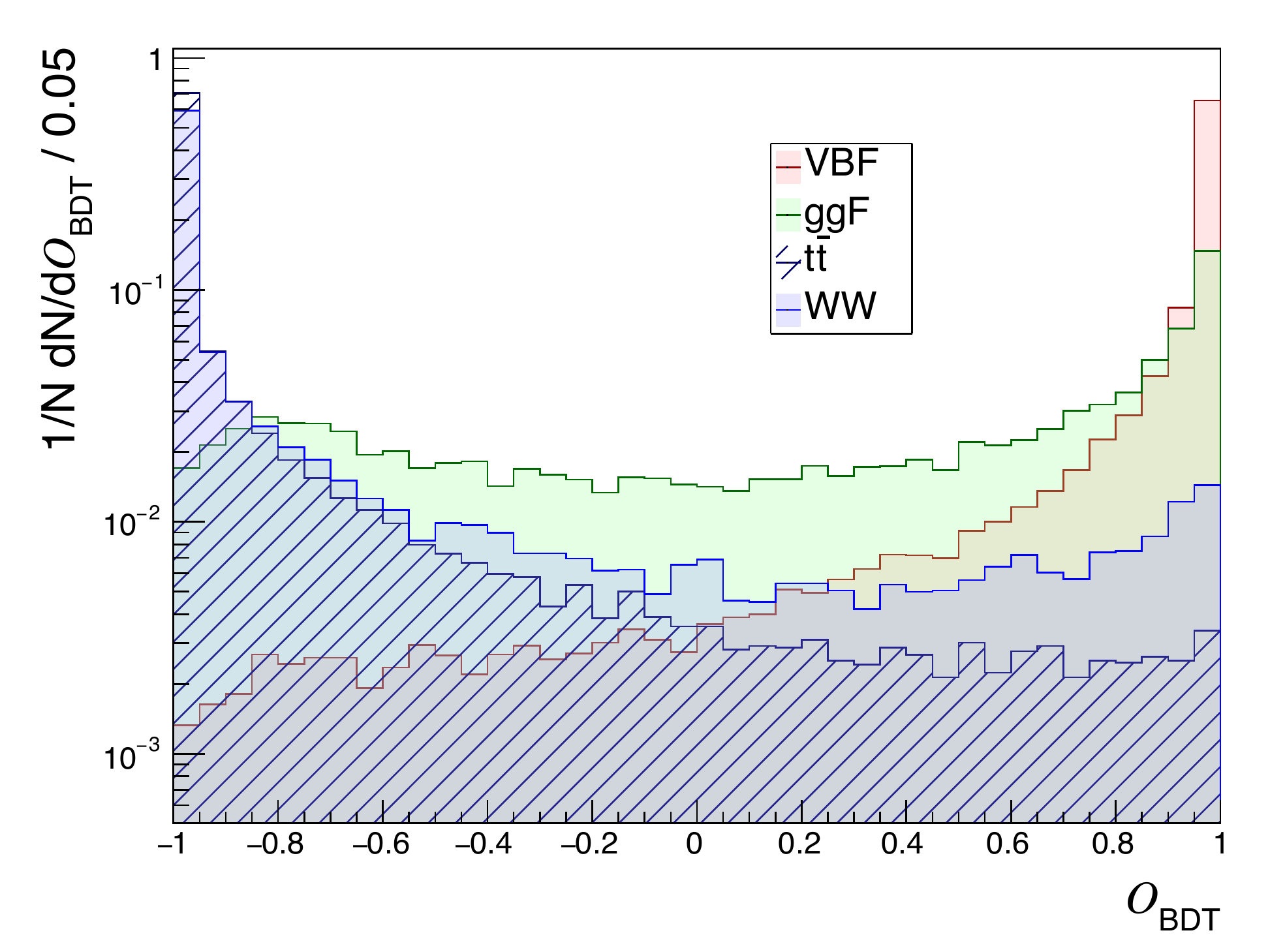}
        \includegraphics[scale=0.4]{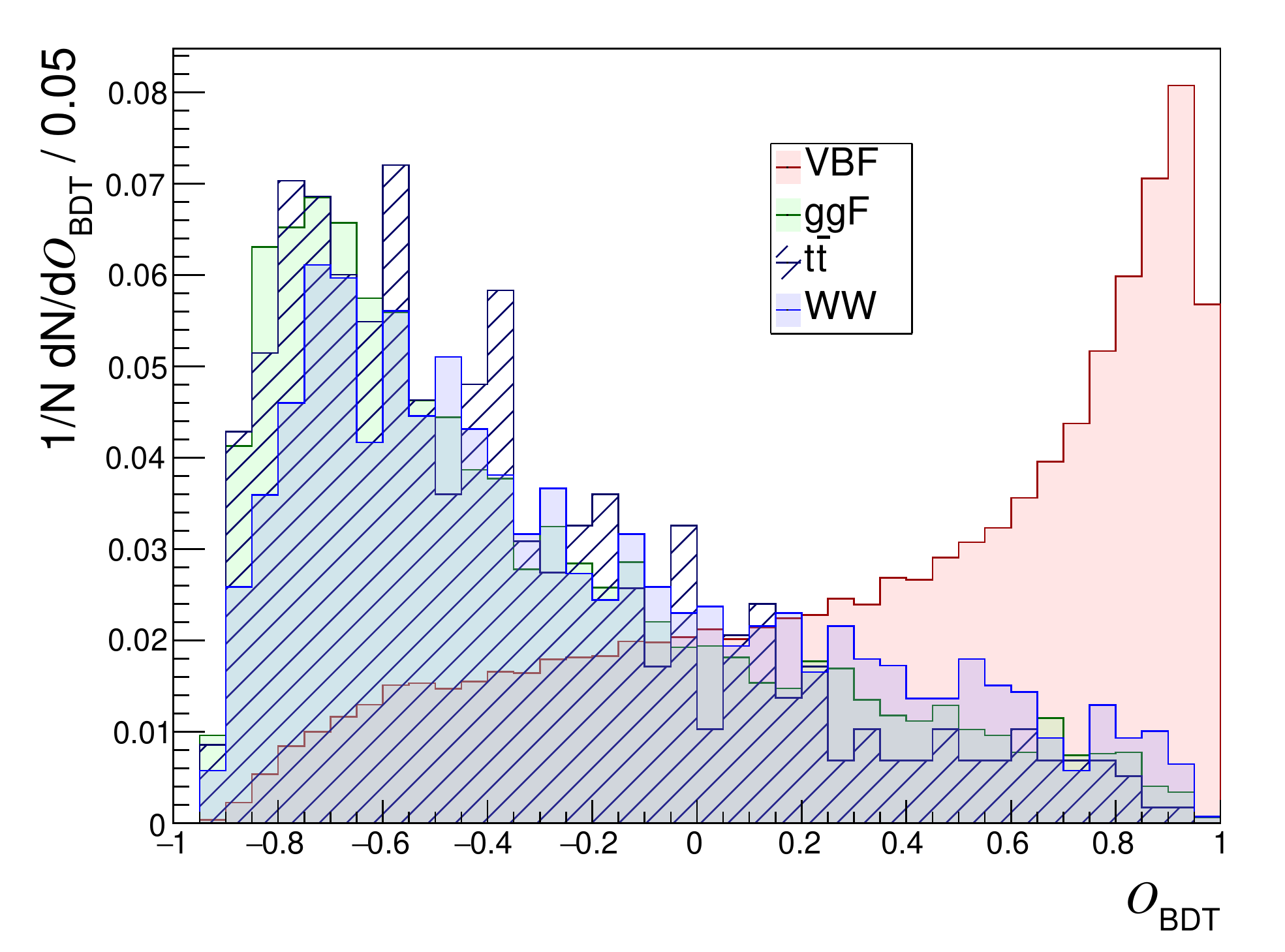}
        \caption{2-step BDT output distributions in first step(left) and second step(right) for each process in $H\to WW^*$.}
	\label{HWW_BDT_All}
\end{figure}

To overcome the problem in the 7-Var BDT, we perform the analysis with a 
new 2-step BDT method.
In the first step, we use the 8 variables as in the standard BDT to
discriminate between the VBF and the SM  backgrounds including $t\bar{t}$ and
$WW$.
Whereas in the second step, we focus on discriminating the VBF and ggF
using the most powerful discriminators as those used in 7-Var BDT. 
Figure \ref{HWW_BDT_All} shows the 2-step BDT output distributions
after both steps. 
The left panel shows the normalized distribution of $O_{\mathrm{BDT}}^1$, in
which near the $-1$ end is more background-like and near the $+1$ end is
more signal-like. Similarly, the right panel shows the 
normalized distribution of $O_{\mathrm{BDT}}^2$ after applying a cut of 
$O_{\mathrm{BDT}}^1 > 0.9$. In a moment, we shall show that the cut value on 
$O_{\mathrm{BDT}}^1 > 0.9$ is the optimal choice with respect to the VBF purity
and ggF contamination.

\begin{figure}[th!]
        \includegraphics[scale=0.5]{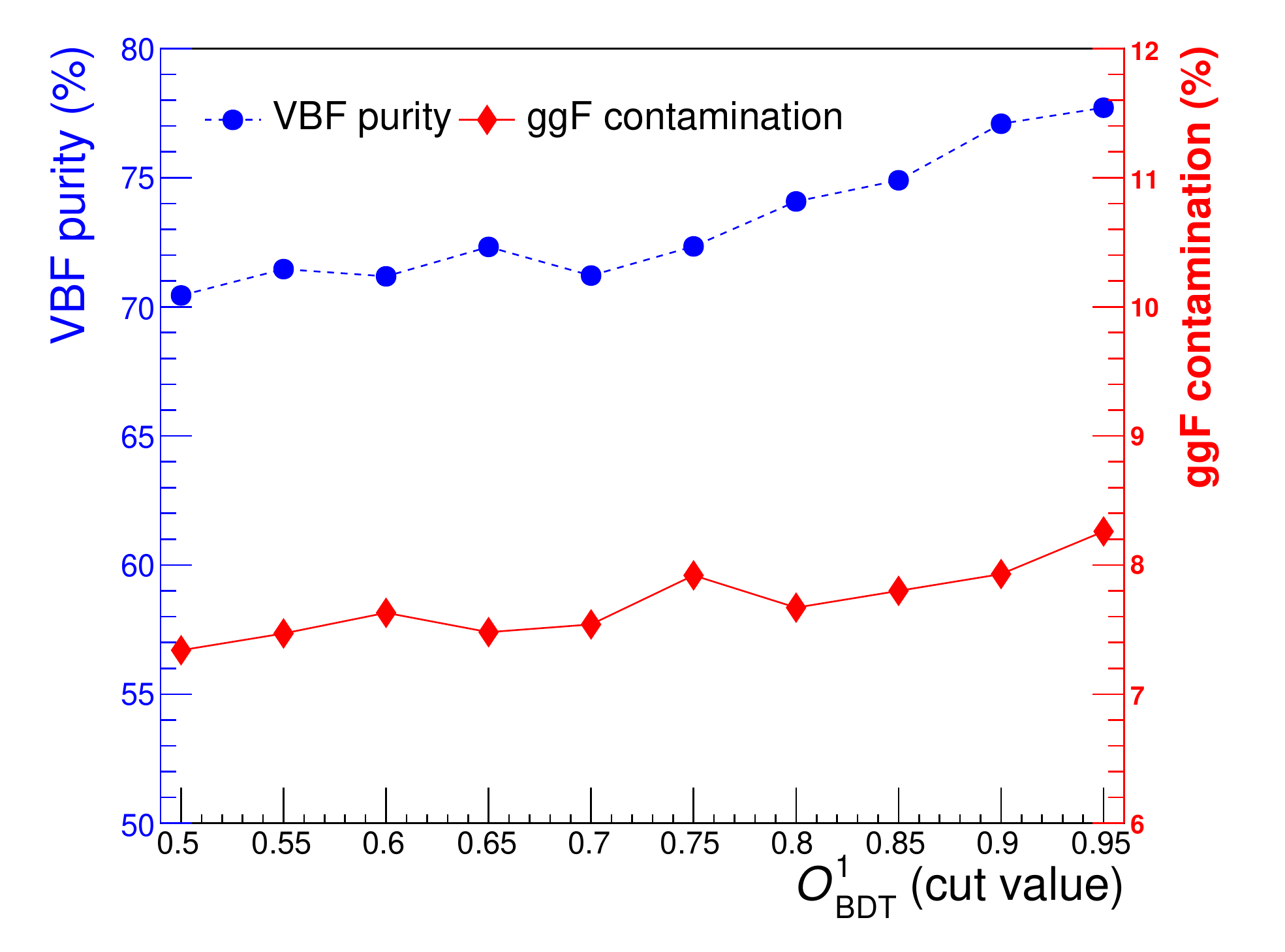}
        \caption{Combined plot of VBF purity and ggF contamination in
 $H\to WW^*$ versus the cut value on $O^1_{\rm BDT}$.}
	\label{HWW_O1}
\end{figure}

Figure \ref{HWW_O1} shows the VBF purity and ggF
contamination versus the cut values of $O_{\mathrm{BDT}}^{1}$ (each
event has a larger value than the cut value). 
It is important to note that the choices of $O_{\mathrm{BDT}}^{1}$ 
and $O_{\mathrm{BDT}}^{2}$ cut values are determined with the signal
efficiency fixed (the signal event number is fixed at 5.1 events 
for various BDT methods).  For example, if $O_{\mathrm{BDT}}^1$ cut 
is set at 0.9 (0.5),
then $O_{\mathrm{BDT}}^2$ cut at 0.166 (0.425), such that the VBF event number is
fixed at 5.1.  Therefore, in Fig.~\ref{HWW_O1} each $O_{\mathrm{BDT}}^1$ 
cut value corresponds to a $O_{\mathrm{BDT}}^2$ cut 
such that the signal event number is fixed at 5.1. 
It is clear and evident that we shall have purer VBF signal sample 
when we impose a more stringent cut.
Also, the ggF contamination increases slightly as the cut gets more 
severe.
The first-step-BDT output cut value is optimized at
$0.9$ to obtain the highest purity of VBF and the lowest ggF
contamination. 
As shown in Table \ref{HWW_result}, with this new method of 2-step BDT
we can highly reduce the ggF contamination down from $12.38$ to $7.93
\%$, and at the same time maintain the VBF purity of 77\%.

 \begin{table*}
    \small
    \begin{tabular}{@{\quad}l@{\quad}llll@{\quad}c@{\quad}c}
    \hline\hline
                 BDT    & \multicolumn{4}{c}{Event number} & VBF purity of & ggF\\
                     \cline{2-5}
                 method  & VBF & ggF & $t\overline{t}$ & $WW$ & all processes & contamination\\
    \hline
    Standard BDT &5.13 &0.73 &0.40&0.45& 76.42\% & 12.38\%\\
    11-Var BDT & 5.11 & 0.61 & 0.32 & 0.43 & 79.05\% & 10.66\%\\
    7-Var BDT & 5.11 & 0.55 & 2.89 & 1.58 & 50.49\% & 9.70\%\\
    2-step BDT ($O_{\mathrm{bdt}}^{1}>0.9)$ &5.10&0.44&0.51&0.56& 77.09\%& 7.93\%\\
    \hline\hline
    \end{tabular}
    \caption{Summary of the results for the event numbers of each
      process, VBF purity, and ggF contamination in $WW$ decay channel,
      after applying cuts on various methods of BDT. Here the ggF
      contamination is defined as
      $N\left(\mathrm{ggF}\right)/\left(N\left(\mathrm{ggF}\right)+N\left(\mathrm{VBF}\right)\right).$
      The event numbers are normalized to 5.8 fb$^{-1}$.
The luminosity here is taken to be the same as in 
Ref.~\cite{atlas-vbf-ww} for direct comparison.
}
    \label{HWW_result}
    \end{table*}

\subsection{$H\to\gamma\gamma$}

\begin{figure}[th!]
        \includegraphics[scale=0.4]{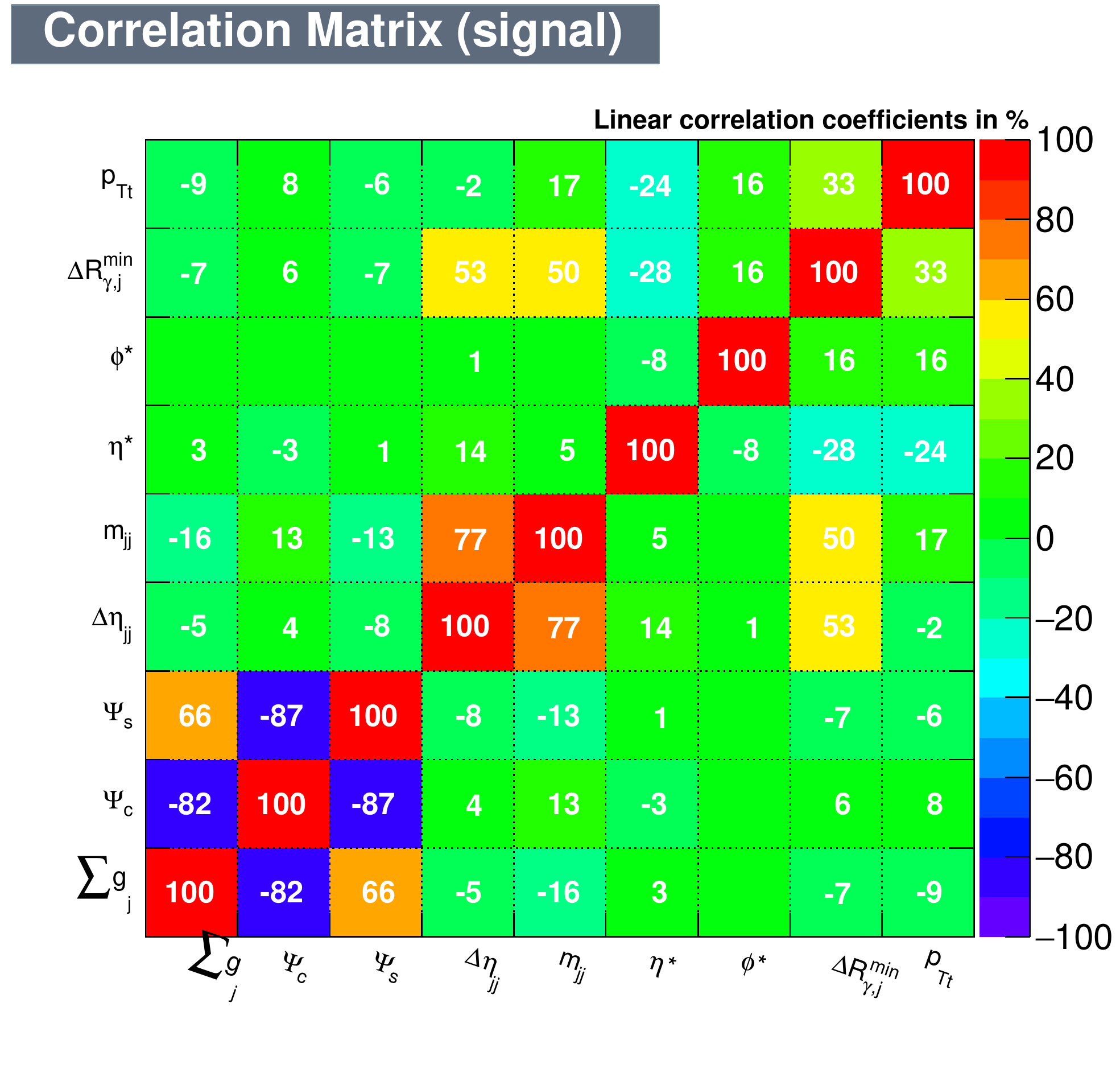}
        \includegraphics[scale=0.4]{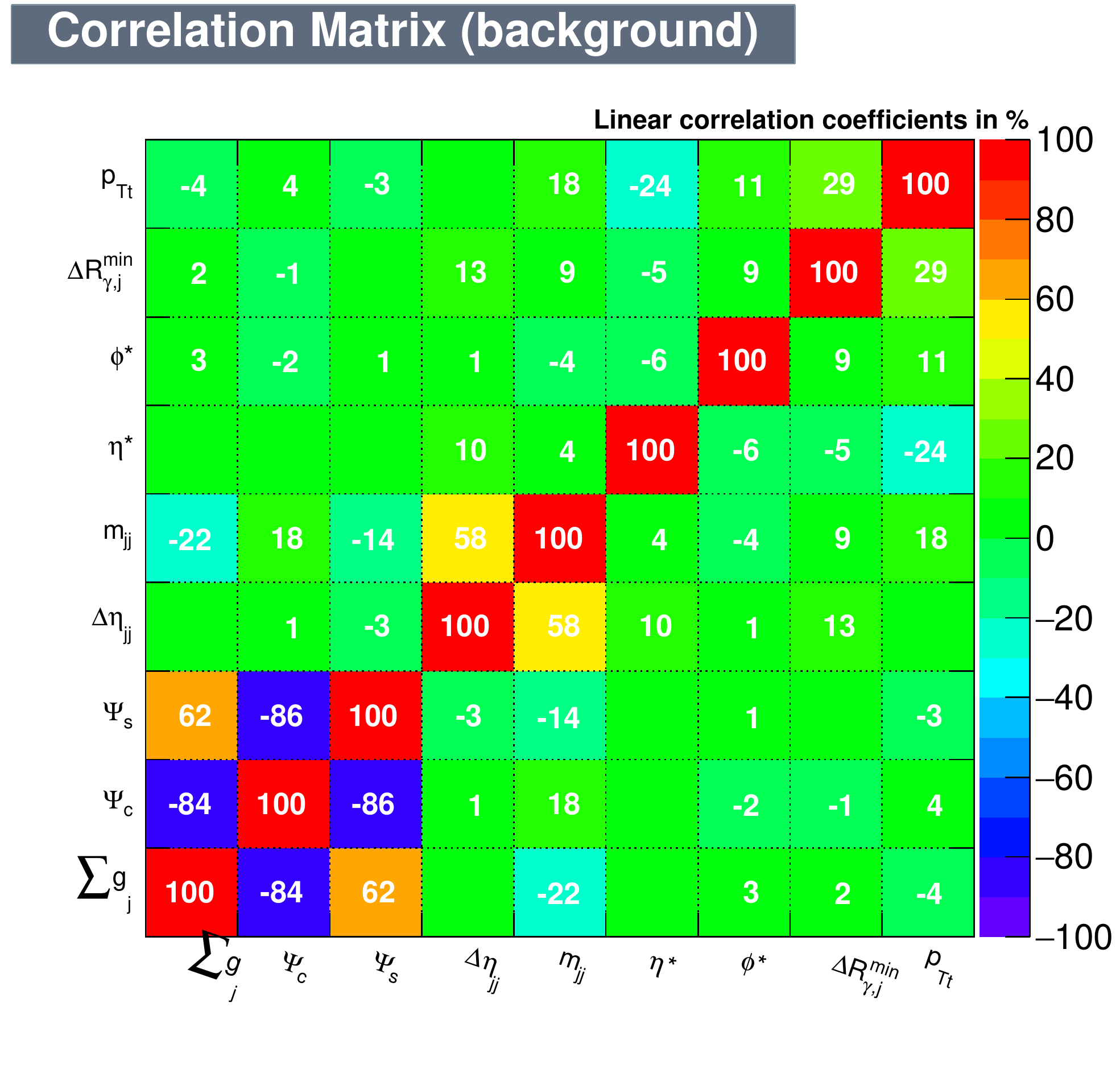}
        \caption{Linear correlation of each variable used in BDT in $H\to\gamma\gamma$. In this figure, the signal (left) denotes VBF, and the 
background (right) includes ggH as well as $\gamma\gamma+jj$.}
	\label{Haa_Corr}
\end{figure}

\begin{figure}[th!]
        \includegraphics[scale=0.4]{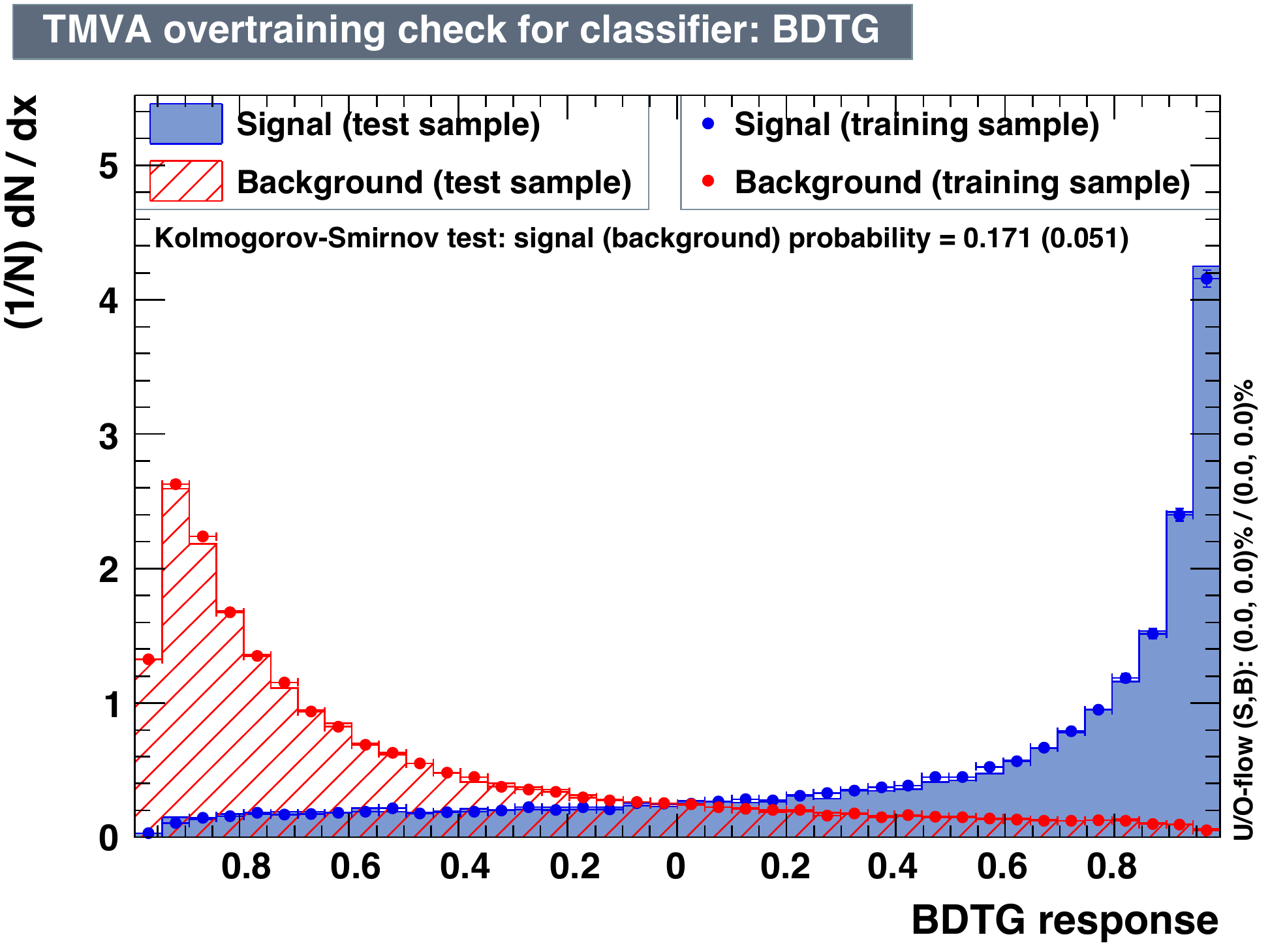}
        \includegraphics[scale=0.4]{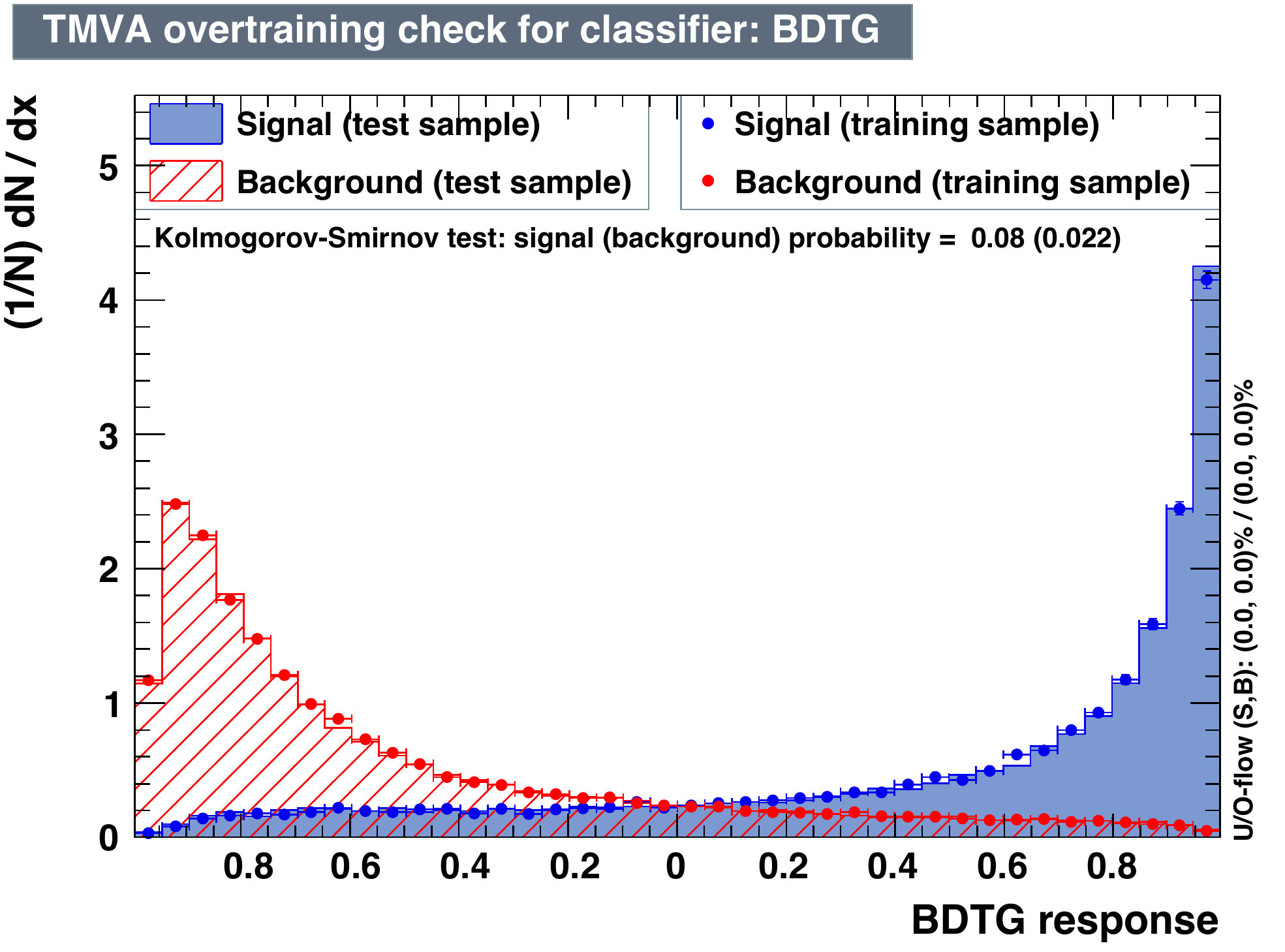}
        \includegraphics[scale=0.4]{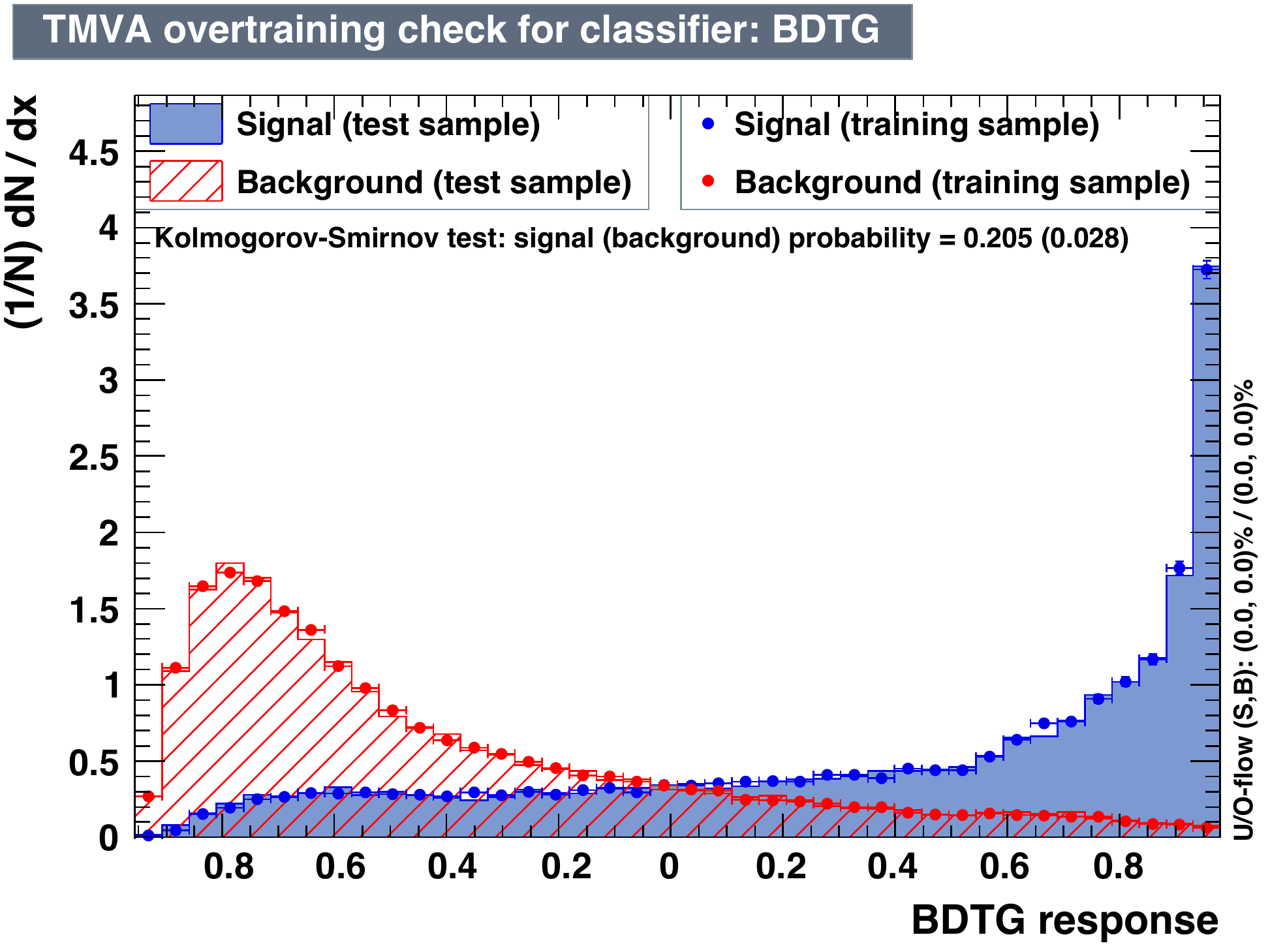}
        \includegraphics[scale=0.4]{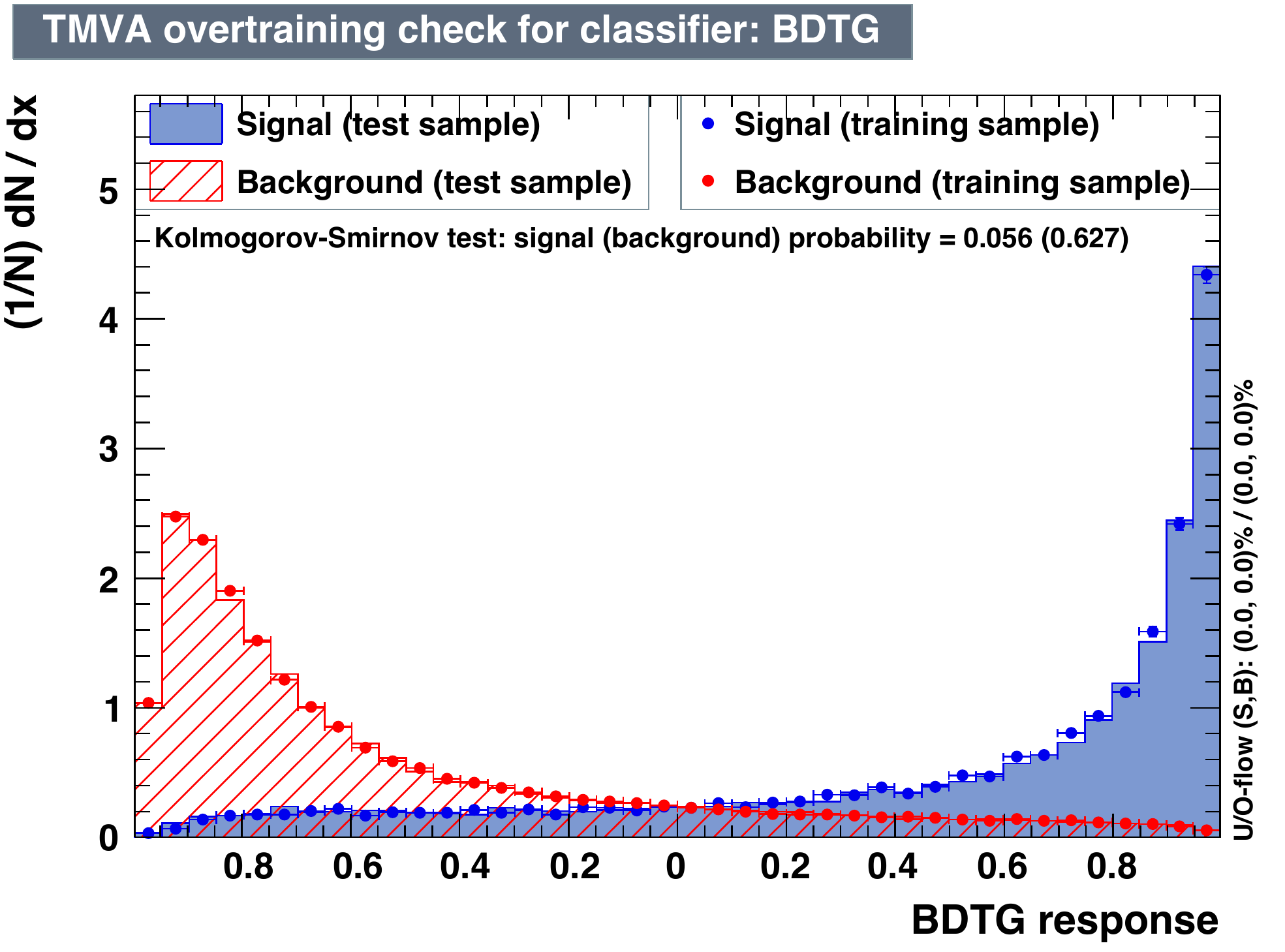}
        \includegraphics[scale=0.4]{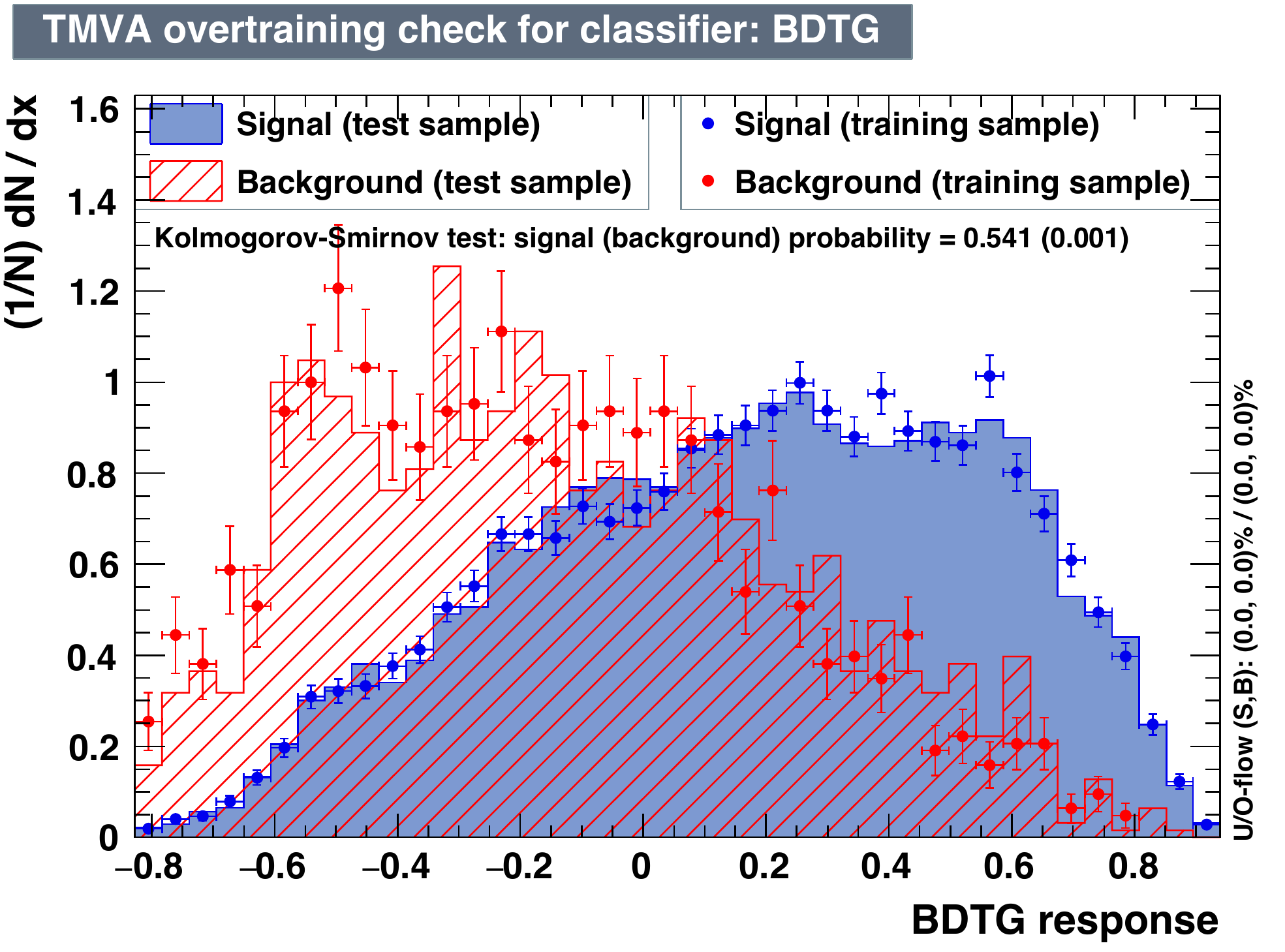}
        \caption{BDT output distribution of training and testing samples
          in $H\to \gamma\gamma$. (\textit{Top-left}) Method of
          standard BDT. (\textit{Top-right}) Method of 9-Var
          BDT. (\textit{Middle-left}) Method of 5-Var
          BDT. (\textit{Middle-right}) The first step in the 2-step
          BDT. (\textit{Bottom}) The second step in the 2-step BDT. 
Note that half of sample is used as the training sample and the 
other half as the testing sample.}
	\label{Haa_BDT}
\end{figure}

In Fig.~\ref{Haa_Corr}, we show the linear
correlations between any two variables that we have used in the channel 
$H\to \gamma\gamma$ analyses. 
We can see that strong correlations among the 3 jet-shape variables, and 
between $\Delta \eta_{jj}$ and $m_{jj}$ in both the signal and background
samples. In addition, in order to avoid overtraining in the BDT analyses, 
we show the BDT output distributions for both the training and testing 
samples as shown in Fig. \ref{Haa_BDT}.

The results of our analyses in the channel $H\to\gamma\gamma$ are 
summarized in Table \ref{Haa_result}. 
We control the VBF efficiency at 5.4\% for
comparison. The 9-Var BDT, which adds 3 new jet-shape variables
compared to the standard BDT, can enhance the VBF purity and at
the same time reduce the ggF contamination by about $2 \%$. 
In order to focus on distinguishing between the VBF and ggF event samples, 
the 5-Var BDT, which uses the most powerful 5 variables, is introduced 
and can further decrease the ggF contamination by about $2 \%.$ 
However, this method sacrifices the discrimination from the other 
SM backgrounds and lowers the 
VBF purity to only 24.6\%.

\begin{figure}[th!]
        \includegraphics[scale=0.4]{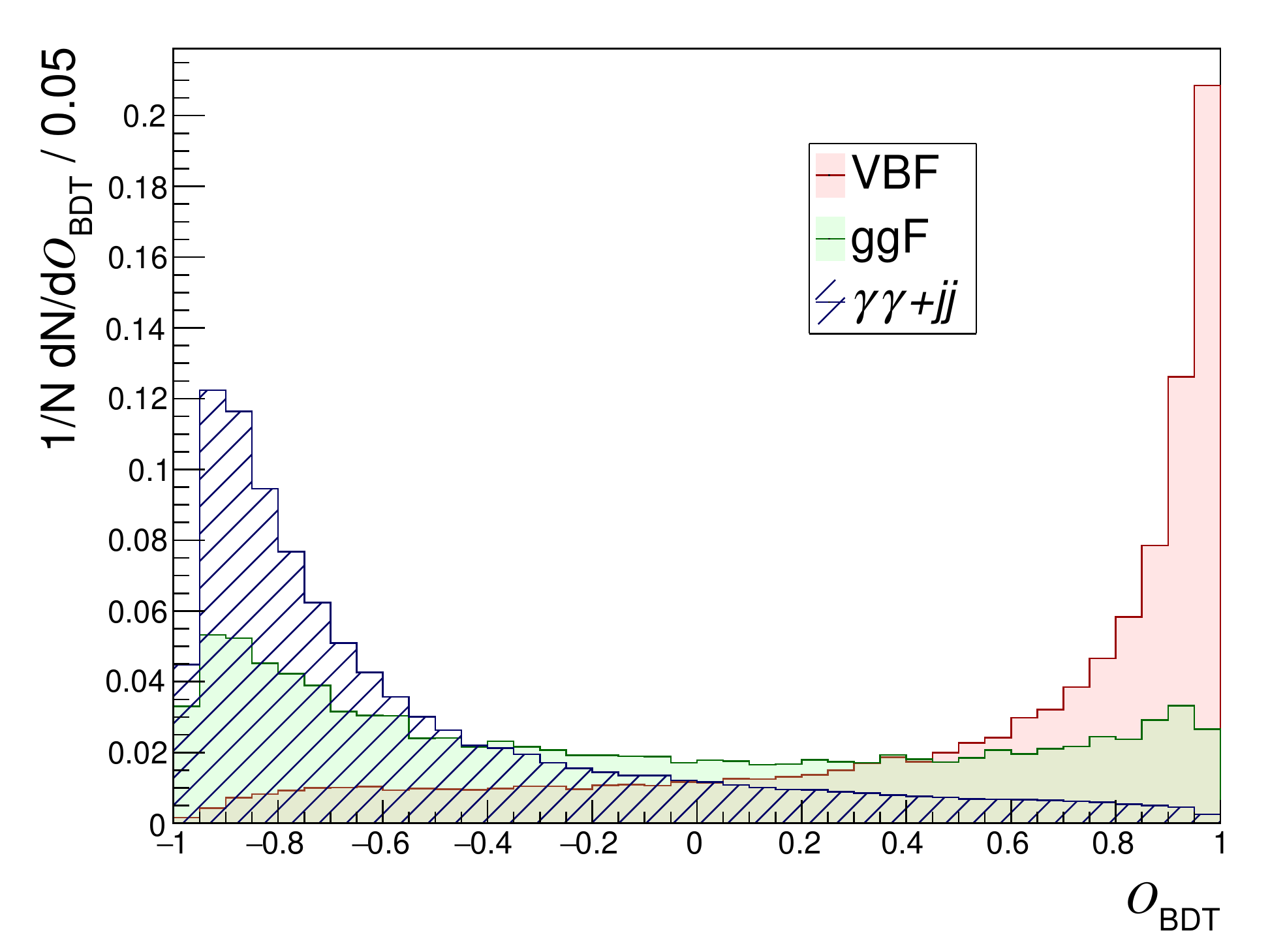}
        \includegraphics[scale=0.4]{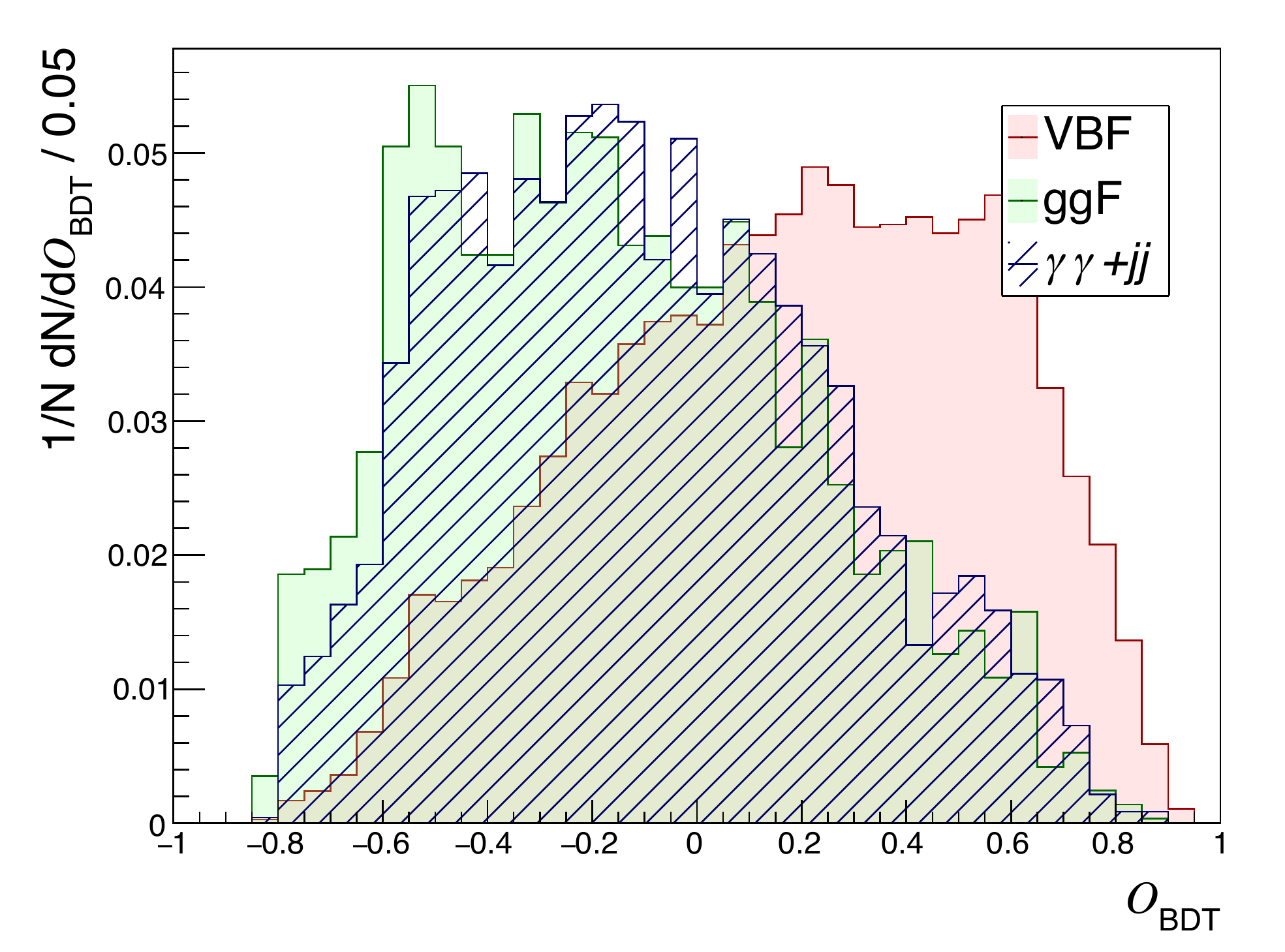}
        \caption{2-step BDT output distributions in first step (left) 
and second step (right) for each process in $H\to \gamma\gamma$.}
	\label{Haa_BDT_All}
\end{figure}

\begin{figure}[th!]
        \includegraphics[scale=0.5]{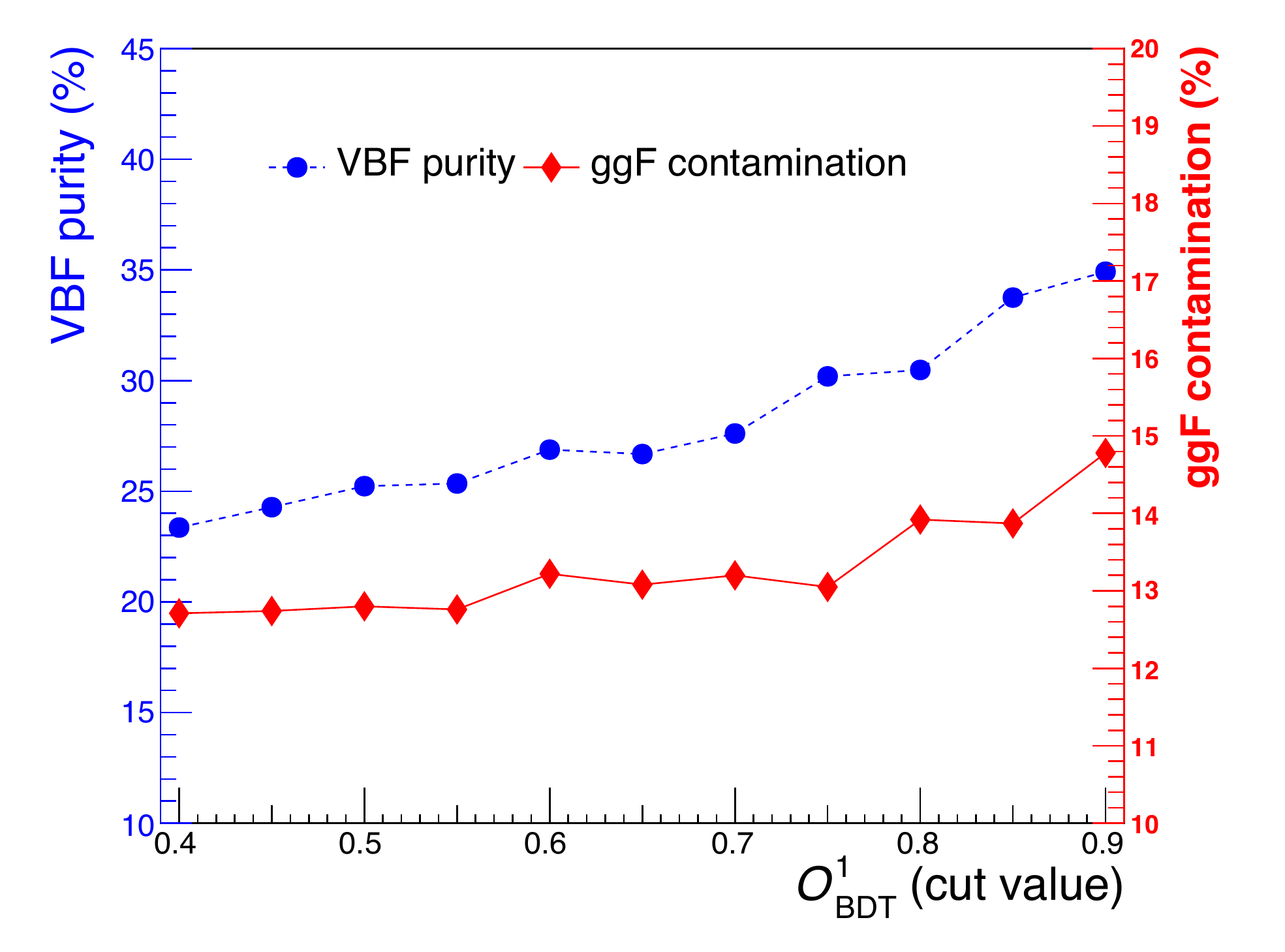}
        \caption{
 Combined plot of VBF purity and ggF contamination in
 $H\to \gamma\gamma$ versus the cut value on $O^1_{\rm BDT}$.
}
	\label{Haa_O1}
\end{figure}

Similar to the previous channel, we attempt the 2-step BDT method
to this case.
We use the standard 6 variables in the first step to
discriminate between the VBF and $\gamma\gamma+jj$ background.
In the second step, we separate between the
VBF and ggF using the most powerful 5 discriminators
as those used in 5-Var BDT. 
Figure \ref{Haa_BDT_All} shows the
2-step BDT output distribution in both steps. 
The left panel shows the normalized distribution of $O_{\mathrm{BDT}}^1$
while the right panel shows the normalized distribution of 
$O_{\mathrm{BDT}}^2$ after applying a cut of 
$O_{\mathrm{BDT}}^1 > 0.75$.
Figure \ref{Haa_O1}
shows the VBF purity and ggF contamination versus the cut value of
$O_{\mathrm{BDT}}^{1}$. 
Similar to the previous channel, the choices of $O_{\mathrm{BDT}}^{1}$ 
and $O_{\mathrm{BDT}}^{2}$ cut values are determined with the signal
efficiency fixed at 5.4\% for various BDT methods.  Therefore, 
each $O_{\mathrm{BDT}}^1$ cut value in Fig.~\ref{Haa_O1}
corresponds to a $O_{\mathrm{BDT}}^2$ cut such that the VBF signal 
efficiency is fixed at 5.4\%.
Again, we can achieve a purer VBF signal sample
but with a slightly larger ggF contamination 
when we apply a more stringent cut value. 
The cut value of $O_{\mathrm{BDT}}^{1}$ is optimized at 
$0.75$ for the highest
purity of VBF and the lowest ggF contamination. 
As shown in Table~\ref{Haa_result}, the ggF contamination is substantially
reduced from $18.89 \%$ to $13.59 \%$, and at the same time 
maintain the VBF purity at about 30.2\%.
\footnote
{The ggF contamination that we obtained by the standard BDT in the channel 
$H\to\gamma\gamma$ is somewhat smaller (about $6\%$) 
than that obtained in ATLAS \cite{atlas-vbf-gg}.
We presume the discrepancy is due to the uncertainty in detector simulations
as we use D{\scriptsize ELPHES} while ATLAS uses G{\scriptsize EANT}4.
}

\begin{table*}
    \small
    \begin{tabular}{@{\quad}l@{\quad}c@{\quad}ccc@{\quad}c@{\quad}c}
    \hline\hline
  BDT    & VBF & \multicolumn{3}{c}{Event number} & VBF purity of & ggF\\
                     \cline{3-5}
                   method  & efficiency & VBF & ggF & $\gamma\gamma+jj$ & all processes & contamination\\
    \hline
 Standard BDT &5.4\% &6.19 &1.44&10.41& 34.3\% &18.89\%\\
 9-Var BDT & 5.4\% & 6.20 & 1.28  & 9.59 & 36.3\% & 17.08\%\\
 5-Var BDT & 5.4\% & 6.19 & 1.12 & 17.86 & 24.6\% & 15.33\%\\
 2-step BDT ($O_{\mathrm{bdt}}^{1}>0.75)$ &5.4\%&6.19&0.97&13.32& 30.2\%& 13.59\%\\
    \hline\hline
    \end{tabular}
    \caption{Summary of the results for the event numbers of each process,
VBF purity, ggF contamination in diphoton decay channel, 
after applying cuts on various methods of BDT. 
The event numbers are normalized to 13.3 fb$^{-1}$.
The luminosity here is taken to be the same as in 
Ref.~\cite{atlas-vbf-gg} for direct comparison.
}
    \label{Haa_result}
    \end{table*}

\subsection{Receiver Operating Characteristic (ROC) curves}

\begin{figure}[th!]
        \includegraphics[scale=0.6]{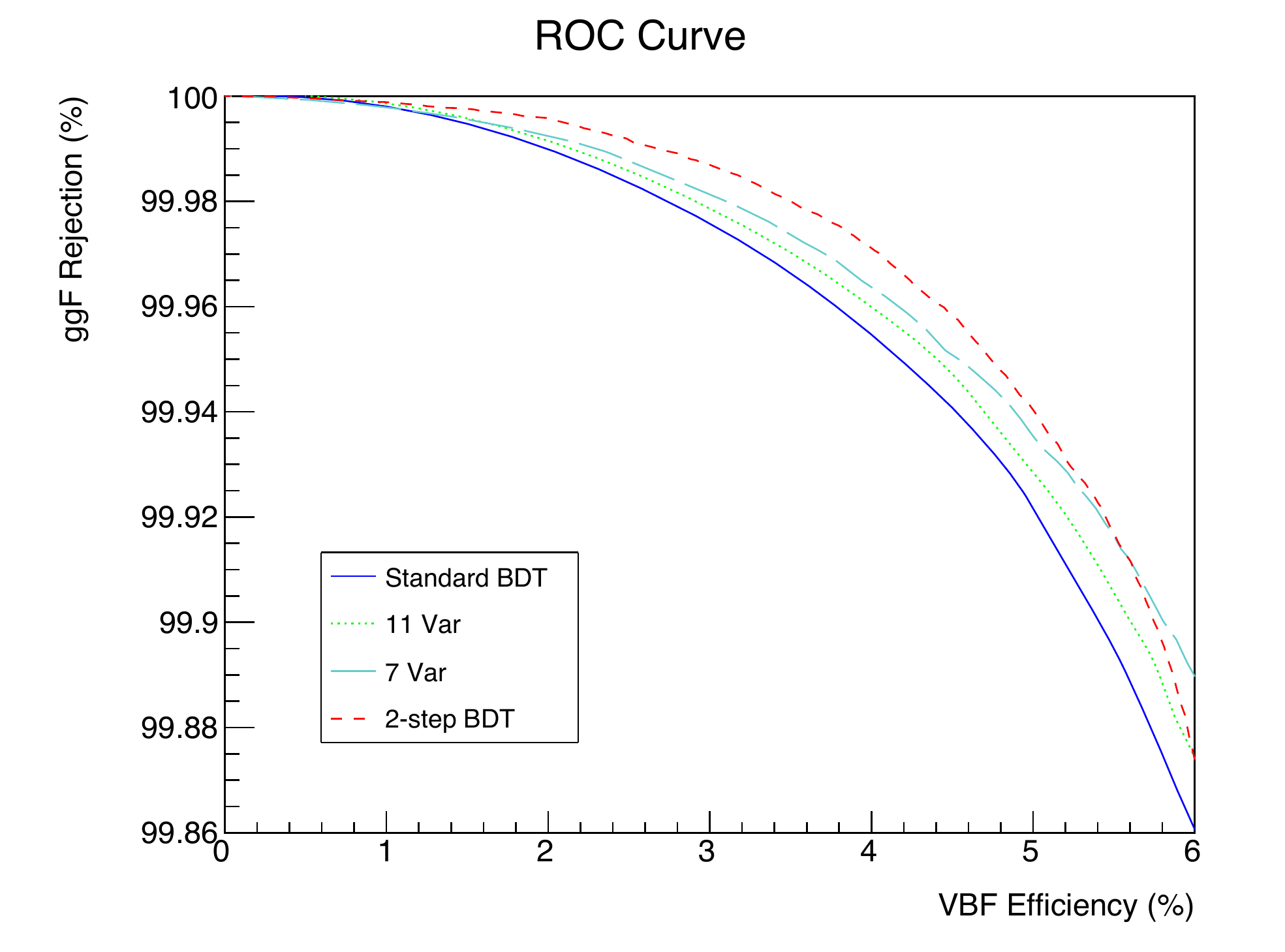}
        \caption{The ROC curves for ggF rejection vs VBF efficiency 
for various BDT methods used in  $H\to WW^*$ channel.  
For the 2-step BDT we have imposed 
$O_{\rm BDT}^1> 0.9$ and then vary $O_{\rm BDT}^2$. 
As indicated in Table~\ref{HWW_result} the final VBF efficiency 
is set at 3.8\% where the event numbers for VBF and ggF in the 
2-step BDT are 5.10 and 0.44, respectively.
}
	\label{roc-ww}
\end{figure}

\begin{figure}[th!]
        \includegraphics[scale=0.6]{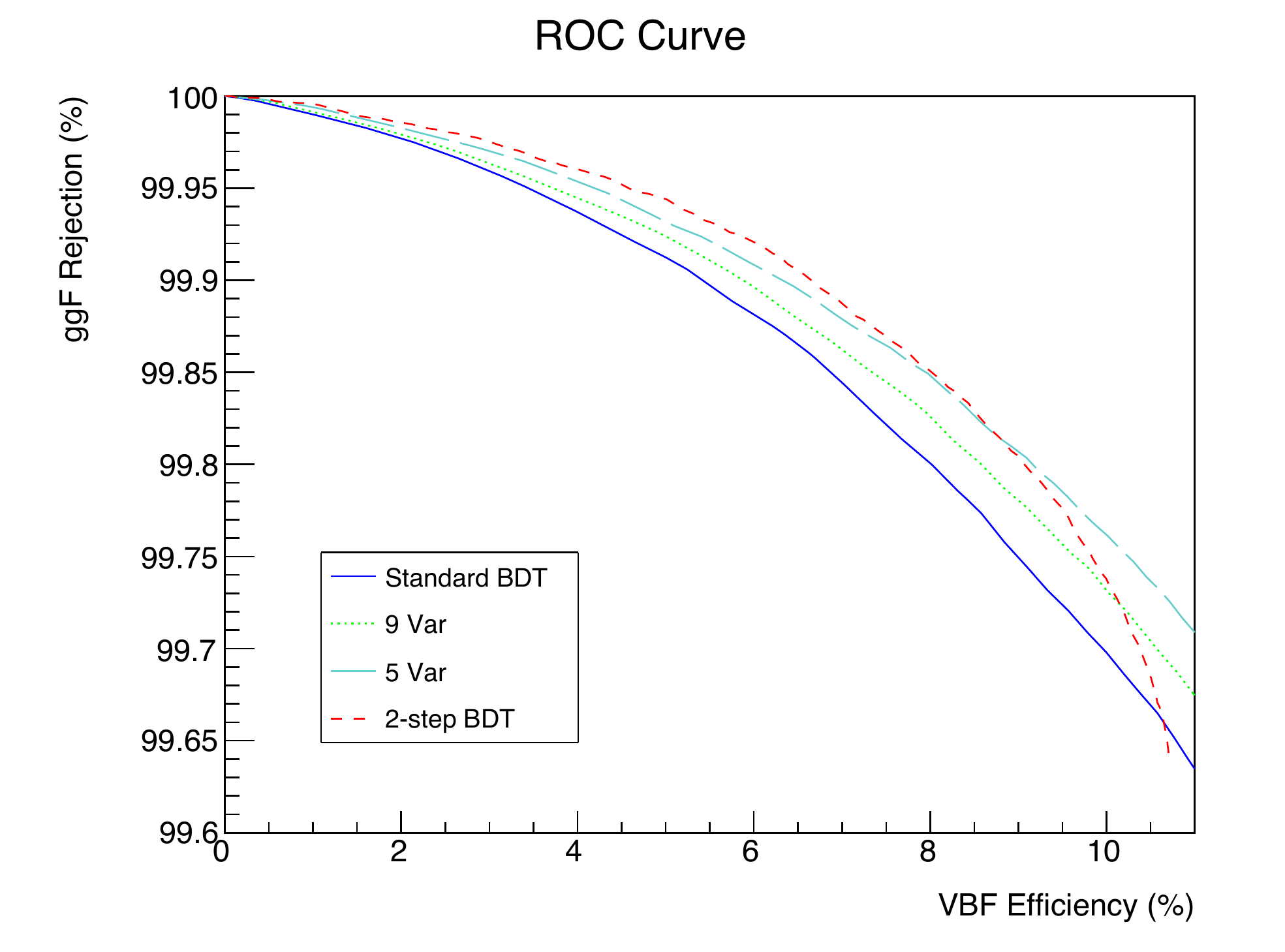}
        \caption{The ROC curves for ggF rejection vs VBF efficiency 
for  various BDT methods used in $H\to \gamma\gamma$ channel.  
For the 2-step BDT we have imposed 
$O_{\rm BDT}^1> 0.75$ and then vary $O_{\rm BDT}^2$. 
As indicated in Table~\ref{Haa_result} the final VBF efficiency 
is set at 5.4\% where the event numbers for VBF and ggF in the 
2-step BDT are 6.19 and 0.97, respectively.
}
	\label{roc-aa}
\end{figure}

Statisitically, it is useful to present the effectiveness of various
methods using the ROC curves, so that one can easily read the effectiveness
of various BDT off the ROC curves.
Here we show parametrically
the gF rejection rate ($y$-axis) versus the VBF efficiency ($x$-axis).
On one side it is the VBF efficiency that we prefer to be large while 
on the other side is the ggF rejection rate that we want to be 
as close to 100\% as possible.  However, in reality the higher VBF efficiency
the lower the ggF rejection will be. 
We show the ROC curves for the $H\to WW$ and $H\to\gamma\gamma$
channels in Fig.~\ref{roc-ww} and Fig.~\ref{roc-aa} , respectively,
where we show the ggF rejection rate vs VBF efficiency.
Note that in the 2-step BDT we have set $O_{\rm BDT}^1 > 0.9 \, (0.75)$ 
for $H \to WW \; (H \to \gamma\gamma)$ channel before
we vary $O_{\rm BDT}^2$ in the figures.
%
In $H\to WW$ channel, the 2-step BDT achieves the best ggF rejection, and thus
the least ggF contamination.
This is consistent with the ggF contamination shown in Table~\ref{HWW_result}.
Similarly, in $H\to\gamma\gamma$ channel, the 2-step BDT offers
the best for ggF rejection.

\section{Conclusions}

We have studied the performance of the approach of 2-step boosted decision
trees. We have followed as closely as the way that the ATLAS generated  
the event samples of VBF, ggF, and the corresponding SM backgrounds in 
the channels of $H \to W W^*$ and $H \to \gamma\gamma$. In the first step,
we trained the VBF signal against the SM backgrounds without the ggF sample,
while in the second step we trained the VBF signal against the ggF sample.

We have demonstrated with our new approach of 2-step BDT, we can
achieve a significant reduction of the ggF contamination
from 12\% (19\%) down to 8\% (12\%) for $H \to W W^*$ ($H\to\gamma\gamma$).
At the same time, we can maintain or slightly improve 
the overall purity of the VBF sample among all the backgrounds. 

The approach of this study can be applied to other decay channels, such 
as $H\to Z Z^*$, $\tau\tau$, and $b\bar b$. 
Further investigations can include 
optimization of the number of variables used in each step 
in the 2-step BDT. Actually, one can use
various ways to  rank the importance of each variable. 

\section*{\bf Acknowledgments}
This research was supported in parts by the MoST of Taiwan
under Grant Nos. MOST-105-2112-M-007-028-MY3 and MOST-103-2112-M-007-024-MY3.


\newpage

\end{document}